\documentclass[preprint,12pt]{aastex}
\usepackage{color}
\usepackage{graphicx}

\slugcomment{}
\shorttitle{}
\shortauthors{}

\newcommand{\dm}{\ensuremath{\Delta m_{\mathrm{15}}}}
\newcommand{\fig}{Figure}
\newcommand{\tab}{Table}

\newcommand{\NSNe}{36}     
\newcommand{\Nfit}{30}     
\newcommand{\Ntemp}{24}    
\newcommand{\Ntempi}{18}   
\newcommand{\NtempNIR}{14}   

\begin{document}
\title{The Carnegie Supernova Project: Light Curve Fitting with SNooPy}
\author{
Christopher R. Burns\altaffilmark{1},
Maximilian Stritzinger\altaffilmark{2,3,4},
M. M. Phillips\altaffilmark{2},
ShiAnne Kattner\altaffilmark{5},
S. E. Persson\altaffilmark{1},
Barry F. Madore\altaffilmark{1},
Wendy L.\ Freedman\altaffilmark{1},
Luis Boldt\altaffilmark{2},
Abdo Campillay\altaffilmark{2},
Carlos Contreras\altaffilmark{6},
Gaston Folatelli\altaffilmark{2,7},
Sergio Gonzalez\altaffilmark{2},
Wojtek Krzeminski\altaffilmark{2},
Nidia Morrell\altaffilmark{2},
Francisco Salgado\altaffilmark{2} and 
Nicholas B. Suntzeff\altaffilmark{8}
}
\altaffiltext{1}{Observatories of the Carnegie Institution for Science, 813 Santa Barbara St, Pasadena, CA, 91101, USA}
\altaffiltext{2}{Carnegie Institution of Washington, Las Campanas Observatory, Colina El Pino, Casilla 601, Chile}
\altaffiltext{3}{Dark Cosmology Center, Niels Bohr Institute, University of Copenhagen, Juliane Maries Vej 30, 2100 Copenhagen \O, Denmark}
\altaffiltext{4}{The Oskar Klein Centre, Department of Astronomy, Stockholm University, AlbaNova, 10691 Stockholm, Sweden}
\altaffiltext{5}{Astronomy Department, San Diego State University, 5500 Campanile Drive, San Diego, CA 92182, USA}
\altaffiltext{6}{Centre for Astrophysics \& Supercomputing, Swinburne University of Technology, P.O. Box 218, Victoria 3122, Australia}
\altaffiltext{7}{Universidad de Chile, Departmento de Astronomia, Casilla 36-D, Santiago, Chile}
\altaffiltext{8}{George P. and Cynthia Woods Mitchell Institute for Fundamental Physics and Astronomy, Texas A\&M University, Department of Physics and Astronomy, College Station, TX, 77843, USA}

\begin{abstract}
In providing an independent measure of the expansion
history of the Universe, the Carnegie Supernova Project (CSP) has
observed 71 high-$z$ Type Ia supernovae (SNe~Ia) in the near-infrared
bands $Y$ and $J$. These can be used to construct rest-frame $i$-band light curves which,
when compared to a low-$z$ sample, yield distance moduli that
are less sensitive to extinction and/or decline-rate corrections than
in the optical.  However, working with NIR observed and
$i$-band rest frame photometry presents unique challenges
and has necessitated the development of a new set of observational
tools in order to reduce and analyze both the low-$z$ and high-$z$
CSP sample. We present in this paper the methods used to generate
$uBVgriYJH$ light-curve templates based on a sample of \Ntemp\ high-quality
low-$z$ CSP SNe. We also present two methods for determining the distances
to the hosts of SN~Ia events. A larger sample of \Nfit\  low-$z$ SNe~Ia in the
Hubble Flow are used to calibrate these methods. We then apply the method
and derive distances to seven galaxies that are so nearby that their
motions are not dominated by the Hubble flow.
\end{abstract}


\section{Introduction}

Type~Ia supernovae (SNe~Ia) are now well established as precise
standard candles. After accounting for the well-known correlation
between peak-magnitude and decline rate $\dm(B)$, the rms
variation from supernova to supernova typically amounts to less than
0.15 magnitudes \citep[hereafter F10]{2010AJ....139..120F}
\citep{1996AJ....112.2391H,2006ApJ...647..501P,2008ARA&A..46..385F,2009ApJ...700.1097H}.
With a typical peak bolometric luminosity of $L_{SN}\simeq10^{43}\mathrm{erg\cdot s^{-1}}$,
SN~Ia can be observed from the ground and space out to cosmological
distances, thereby constraining the expansion history of the Universe
\citep{2009ApJ...704.1036F,2009ApJS..185...32K,2007ApJ...666..694W,2006A&A...447...31A,
1999ApJ...517..565P,1998AJ....116.1009R}.
They can also be used in the local Universe to determine distances
to galaxies that are beyond the reach of more accurate distance indicators
such as Cepheid variables, yet are close enough that large scale structures
could significantly perturb the Hubble distance.

It is well known that the decline-rate corrections for SNe~Ia are largest
in the ultra-violet bands (where the correction can be as high as
0.5 mag), decrease steadily through the optical bands, and are almost
non-existent in the NIR
bands 
\citep{2004ApJ...602L..81K,2008ApJ...689..377W}.
Furthermore, SNe~Ia, like all standard candles, are affected by interstellar
extinction both from the Milky Way and their host galaxies, to say
nothing of any possible extinction in the intergalactic medium (IGM).
Extinction by dust is known to decrease with wavelength
\citep[hereafter CCM+O]{1989ApJ...345..245C,1994ApJ...422..158O}.  For a
typical line of sight in the Milky Way (where $R_V\sim3.1$) with 1 magnitude of
extinction in $U$ band, we would expect extinctions of 0.85 mag in $B$ band,
0.66 mag in $V$ band, 0.54 mag in $R$ band, 0.39 mag in $I$ band, 0.19 mag
in $J$ band, 0.12 mag in $H$ band, and 0.08 mag in $K$ band.
For these reasons, the CSP measured the expansion history of the Universe
by constructing a rest-frame $i$-band Hubble diagram, thereby reducing
its exposure to decline-rate calibration uncertainties and 
interstellar extinction corrections.
In so doing, the CSP was designed to provide an independent constraint
on the cosmology which is less sensitive to a variety of
different systematic errors
\citep{2009ApJ...704.1036F,2010AJ....139..120F}.

In addition to using an independent high-$z$ data set, the CSP
also has the advantage of an independent, high-quality sample of low-$z$
SNe~Ia that have been observed in several wavelengths: $uBVgriYJHK_s$,
all using the same telescope and set of filters at the Las Campanas
Observatory (LCO). This sample of approximately 100 SNe~Ia will provide a
uniform sample in a photometric system that is well understood and
will be invaluable for future supernova studies \citep{2010AJ....139..519C}. 

SNe~Ia are transient objects whose light curves rise
very rapidly. As a consequence, it is often the case that dedicated
observations begin only after the peak has occurred. Furthermore, using
SNe~Ia for cosmological experiments requires working with light curves
whose signal-to-noise ratio (S/N) is significantly lower than the
low-$z$ sample to which they are compared. For these reasons,
there is a need to fit the observed light curves with interpolating
functions, often termed light-curve templates, constructed from well-sampled,
high S/N light curves of nearby SNe~Ia. It is well known that there
is a significant evolution of the light-curve shape with decline rate,
especially at longer wavelengths, and the light-curve templates should
capture this behavior. The resulting light-curve templates can then
be fit to the observations, resulting in an estimate of the time of
maximum, peak magnitudes, and decline rate. By combining multi-band
photometry, one can also learn about the amount of reddening, the
reddening law, and discriminate amongst SN~Ia sub-types.

\begin{figure}
\includegraphics[width=3in]{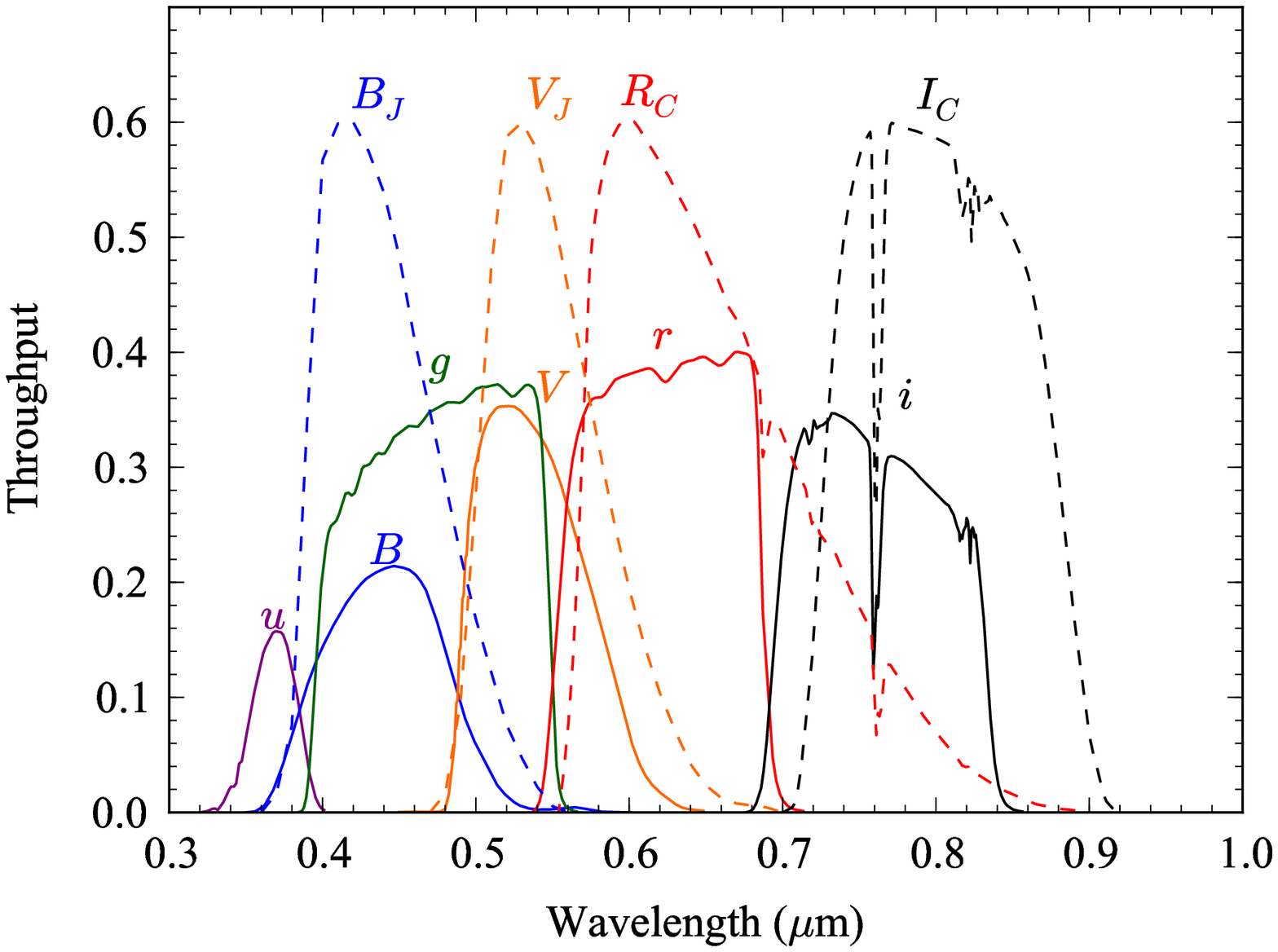}\includegraphics[width=3in]{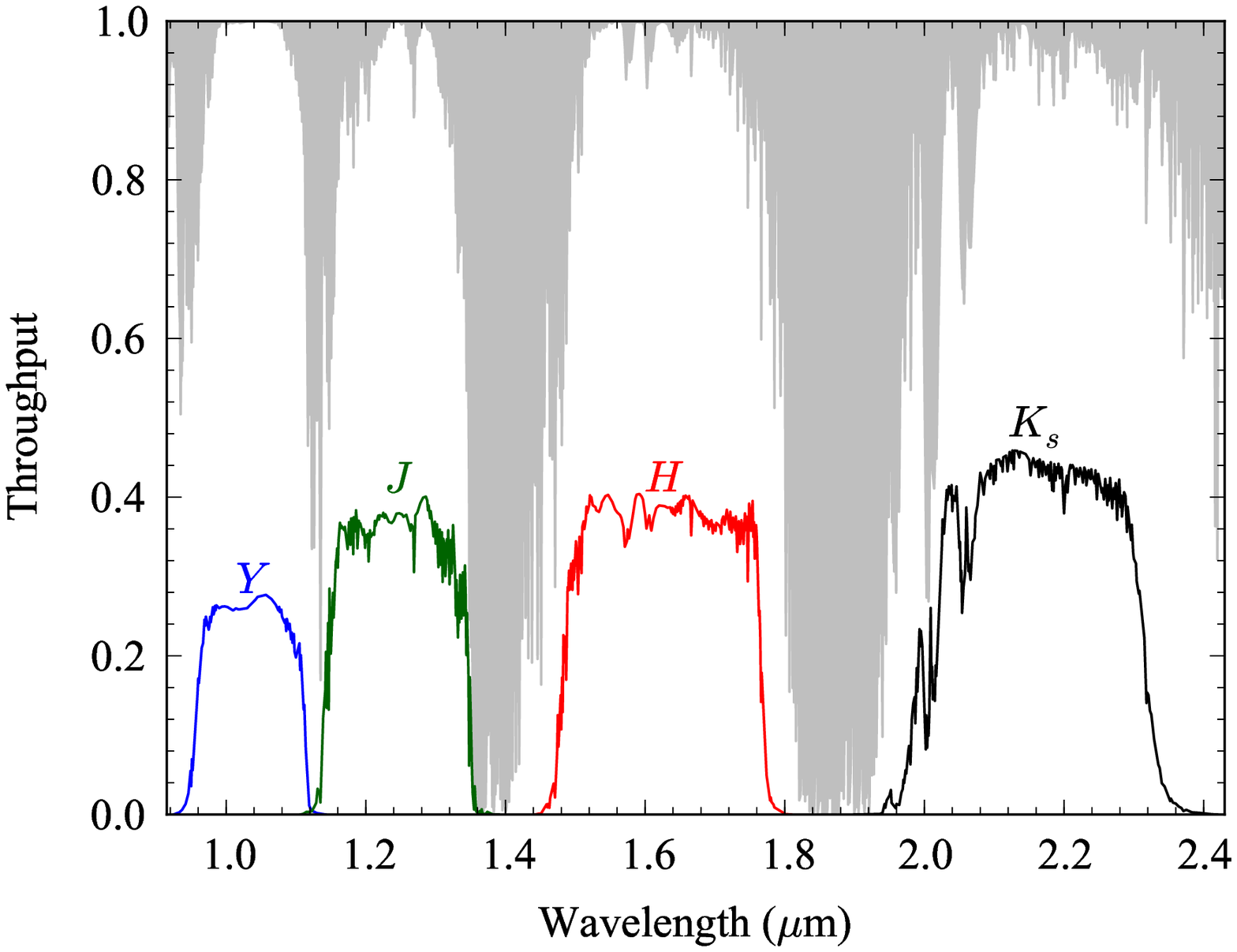}

\caption{The CSP filter band-passes. Left: the optical filter set $uBVgri$ are
plotted as solid lines while the Johnson $BV$ and Cousins $RI$ are
plotted as dashed lines for comparison \citep{2005PASP..117..810S}.  Right: the NIR filter set $YJHK_s$
are plotted as solid lines and the atmospheric absorption at LCO is
plotted as the gray region.\label{fig:CSP-filters}}
\end{figure}

There are several well-established light-curve fitting methods in
the literature. At the time the CSP began to analyze the data from
its first campaign, the two leading methods were SALT \citep{2005A&A...443..781G}
and MLCS2k2 \citep{1996ApJ...473...88R,2007ApJ...659..122J}.
SALT, which generates light-curve templates by modeling the 
underlying SN~Ia spectral energy distribution (SED), could not
be readily used as its SED wavelength range 
did not include
rest-frame Sloan $i$ band. MLCS2k2 was capable of fitting rest-frame 
Johnson $I$ band,
however that would have required the use of non-trivial 
transformations -- termed S-corrections \citep{2002AJ....124.2100S} --
from our low-$z$ Sloan $i$-band observations to Johnson $I$ band.
The transmission functions of $i$ band and $I$ band are significantly 
different (see \fig\ \ref{fig:CSP-filters})
and computing S-corrections would have introduced a significant source
of error. Presently, SALT2 \citep{2007A&A...466...11G} and SifTO
\citep{2008ApJ...681..482C} have joined the ranks of light-curve fitters and
are capable of working in rest-frame $i$. However, these software packages
are all optimized to work at optical wavelengths and in some cases
are ``trained'' using significantly different passbands than the
CSP. It was therefore decided that the CSP would generate its own
light-curve templates based on its $uBVgriYJHK_s$ natural system, with
emphasis on generating accurate light curves in the NIR wavelengths.
We also wanted software that would be easy to use, and also easy to
extend by adding new SNe~Ia and new filters to the sample.
Python\footnote{http://www.python.org}
was chosen as the software development environment because
of its portability to most operating systems, the availability of
powerful numerical and astronomical modules, and its open-source license.
The simple light-curve generating code soon evolved into a more general
package for the analysis of SN~Ia light curves and spectra,
called SNooPy%
\footnote{Following the well-established convention of ending (or beginning)
Python-based packages with ``py'', SNPy was a logical acronym.
Adding the ``oo'' removed any ambiguity in pronunciation (and can
stand for ``object-oriented''). We have since discovered the existence
of a photometry package called SNOoPY. It is our hope the difference
in case will avoid confusion.  The SNooPy software package is available
from the CSP website:  http://www.obs.carnegiescience.edu/CSP%
}. 

This paper outlines the numerical methods used by SNooPy to generate
template light curves in the CSP natural system passbands and the models
used to derive distances to SNe~Ia. Section \ref{sec:CSP-photo}
briefly describes the
CSP photometric system. We describe the methods for generating light-curve
templates and estimating the statistical and systematic errors therein
in section \ref{sec:templates}. 
Section \ref{sec:distances} discusses the use of templates in determining
distances in both the local Universe and for cosmology. Section 
\ref{sec:conclusions}
presents a summary and forecasts future work.

\section{The CSP Photometric system}
\label{sec:CSP-photo}
The low-$z$ CSP is strictly a follow-up program, relying on other projects
to provide us with SN events. Our primary source was the Lick Observatory 
Supernova Search \citep[LOSS]{2001ASPC..246..121F}. We also observed events from 
other surveys including the SDSS-II Supernova
Survey \citep{2008ARA&A..46..385F}, the Catalina Sky Survey \citep{2009ApJ...696..870D},
the CHilean Automatic Supernova sEarch \citep[CHASE]{2009RMxAC..35R.317P},
as well as from many amateur observers%
\footnote{A complete list of SNe, including discovery credits, can be found
on the CSP website: http://www.obs.carnegiescience.edu/CSP%
}.  \tab~\ref{tab:SNe} lists the names and properties of the \NSNe\
SNe~Ia observed by the CSP that
will be used in this paper to generate light-curve templates and calibrate
the distance methods.  This sample comprises the 34 SNe~Ia from 
F10 as well
as 2 other SNe~Ia that were added in order to improve our NIR light-curve
templates.  The photometry of these 2 additional SNe~Ia (SN~2006et and
SN~2007af) will be presented in the next CSP data release paper (Stritzinger et al. in preparation).  
Included in Table~\ref{tab:SNe} are the host galaxy names, 
recessional velocities (heliocentric and CMB-frame), 
time of earliest photometric observation of the SN, and Milky Way reddening.

As well as photometric observations, the CSP has observed most of
its candidates spectroscopically. Aside from determining the type
of SN, spectral coverage allows us to compute more accurate K- and
S-corrections, as well as contribute to the growing library used to
create composite spectra of SNe~Ia \citep{2007ApJ...663.1187H,2002PASP..114..803N}. 

The CSP began observations in 2003, using the Swope 1-m telescope and
du Pont 2.5-m telescope at LCO. The Swope's direct
CCD was used to obtain the optical photometry ($uBVgri$) of SN events
and its NIR camera (RetroCam) was used to obtain NIR photometry in
$YJH$ for the brightest events. The du Pont was used to obtain host galaxy
observations after the SNe had faded using both the direct CCD and
the NIR camera (WIRC). WIRC was also used to obtain NIR observations
in the $YJHK_s$ bands. The details of the observations, data reduction, and 
photometric systems are described in detail in 
\citet{2006PASP..118....2H}, \citet{2010AJ....139..519C},
and \citet{2010AJ....139..120F}.
For this paper, we wish to point out that all  CSP photometric data are
presented on the natural system of the Swope. That is to say, the photometric data
points have been calibrated to zero-points defined by the CSP band-pass
responses on a standard SED (in our case, Vega). The band-passes
are constructed from the filter transmission functions, telescope
and CCD efficiencies, and estimates of the atmospheric absorption.
They have \emph{not} been transformed to some idealized standard filter
system (e.g., Johnson $BV$, Cousins $RI$, etc). This greatly simplifies
the use of our data by other groups, as S-corrections are much more
straightforward (see appendix \ref{sec:CSP-Photometry}).

The definition of the CSP photometric system relies entirely on the
transmission functions and the standard SED. The latter has been measured
to a high degree of precision by \citet{2007ASPC..364..315B} using
the {\em Hubble Space Telescope}. The CSP passbands (see \fig\ \ref{fig:CSP-filters}), 
on the other hand, are
constructed from manufacturer's specifications, models of reflectivity/transmissivity
of the optical components, and measurements of the atmospheric absorption.
These proved inaccurate, as they did not correctly predict the observed
color terms measured at the telescope. We therefore modified the theoretical
response curves by shifting them in wavelength until they predicted
the correct color terms. For $u$ band, we also had to cut the blue-end
of the filter. The details of these corrections can be found in
Appendix A of \citet{2010AJ....139..519C}.  To improve this situation, a group from Texas A\&M
have used a monochromator and photo-diode setup to measure the transmission
functions directly. Preliminary results indicate that, except for the
$u$ band, our theoretical curves are very close to the actual curves. The updated
transmission functions and zero-points will be presented in the next CSP SN~Ia 
data release.

\section{Template Generation}
\label{sec:templates}
At present, SNe~Ia are most often analyzed in terms of their light curves.
In the blue bands, a typical SN~Ia rises rapidly after the initial
explosion, reaching a peak approximately 19 days later and then decaying
on a time-scale of a few months. This morphology naturally introduces
several general characteristics: a peak brightness, the time of the
peak, and the ``width'' of the light curve. Observing in $N$ different
wavelength bands also allows a determination of $N-1$ independent
colors. If SNe~Ia are truly standardizable candles, then any point
on the light curve is as good as any other, though the peak is a natural
point on which to focus, since the S/N is highest there
and its temporal location is unambiguous. As such, many of the light-curve
parameters are defined in terms of the peak. However, models are usually
fit to the entire light curve, and so all the data points contribute (in a weighted
fashion) to the estimation of the parameters. So even though a SN
may not be observed at maximum, its peak brightness can still be inferred
from the rest of the light curve. In this section, we detail how we
parameterize the SN~Ia light curves, how we classify each SN in our
sample, apply K-corrections, and finally construct the light-curve
templates.

\subsection{Parameterization}
\label{sec:param}
Early on, it was determined that the $B$, and to some extent $V$ band
light curves could be described by a single light-curve template that
one would ``stretch'' in the time domain to fit the observed data
\citep{1999ApJ...517..565P}. Yet, as one moves to the longer wavelengths,
an inflection develops around day 20 in $r$ band and evolves into a secondary
maximum when observed in the $i$ band. This second peak increases
in prominence to the point where it can be the brighter of the two
peaks in the NIR bands. It also becomes evident that a simple ``stretch''
correction cannot account for the more complicated morphology, given
that as the stretch decreases, the NIR secondary peaks become
progressively weaker.

Instead of using a stretch-like correction, we can assume that a light-curve
template for any particular filter is a 1-parameter family of curves.
The different light-curve fitting packages in the literature
all use different parameters to describe the shape: MLCS2k2 uses
a luminosity parameter $\Delta$ \citep{1996ApJ...473...88R,2007ApJ...659..122J},
while SALT2 uses a stretch-like parameter $x_{1}$ to describe a sequence of
SEDs of SN~Ia further modified by a color-like parameter $c$ \citep{2007A&A...466...11G}.
SNooPy is a natural extension to the method of \citet{2006ApJ...647..501P},
which built on earlier work by \citet{1996AJ....112.2438H} and so
we choose to use the decline-rate parameter first introduced by \citet{1993ApJ...413L.105P},
$\dm(B)$, defined as the change in magnitudes between the
peak and day 15 of the rest-frame $B$-band light curve. The advantage
of this parameter lies in its simplicity: it is a characteristic of
the observed light curve and is easily measured. Its disadvantage
is that it is tied to a particular filter and photometric system (in
our case, the CSP natural $B$ filter) and is therefore not universal.
It is also based on two specific epochs that may not be observed and
so some degree of interpolation is required to measure it.

Our first task, then, is to measure $\dm\left(B\right)$
for all the SNe~Ia we can in our sample. By necessity, these will
be SNe for which a clear and unambiguous peak is observed in the $B$
light curve. Column 5 of \tab\ \ref{tab:SNe} shows the epoch of the
first observation of each of our SNe, relative to $B$-maximum.  Those
SNe for which $t_0 < 0$ can be used for creating $B$ templates.  
The light curves for other filters peak at different times relative to $B$, 
so not all
filters will necessarily have a well-defined peak, particularly the
NIR bands, where the peak is typically 4 days prior to $B$ maximum.   
Column 7 of \tab\ \ref{tab:SNe} indicates the filters whose
light curves have well-defined
peaks and are 
suitable for creating templates.
However, before we can go about measuring $\dm\left(B\right)$, and
determining peak magnitudes,
we need to correct for the red-shifting of each SN~Ia's SED.

\subsection{K-Corrections}

The first step in the process is to correct for the fact that the
observed supernova SED has been redshifted by an amount $(1+z_{hel})$
and so we are effectively observing with filters that have been blue-shifted
in the rest-frame of the SN. We must therefore K-correct the observed
photometry. The procedure we adopt for doing this is effectively the same
as that used by \citet{2007ApJ...663.1187H}, which we shall briefly
outline here.

\begin{figure}
\includegraphics[width=4in]{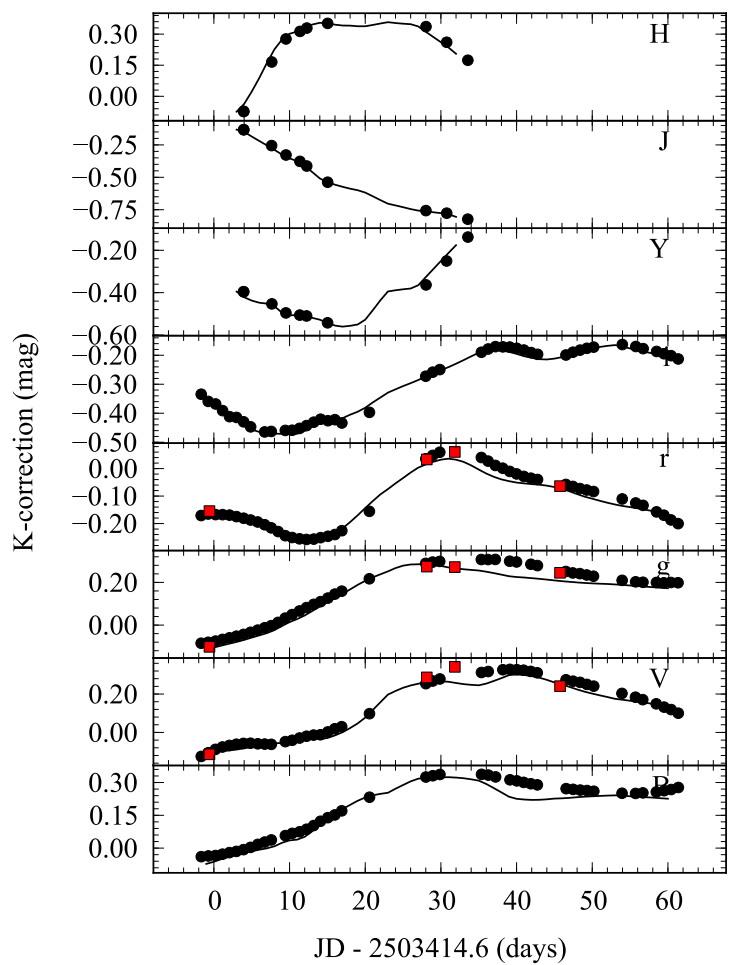}

\caption{Example K-corrections for SN~2005ag, one of the highest redshift 
SN~Ia in our sample. The K-corrections are computed in 3 ways: (1) using
the unmodified \citet{2007ApJ...663.1187H} SED (lines), (2) warping
the \citet{2007ApJ...663.1187H} SED to match the observed colors
(round points), and (3) using the observed spectrum of SN~2005ag (red
squares). \label{fig:K-corrections}}
\end{figure}

\begin{figure}
\includegraphics[width=4in]{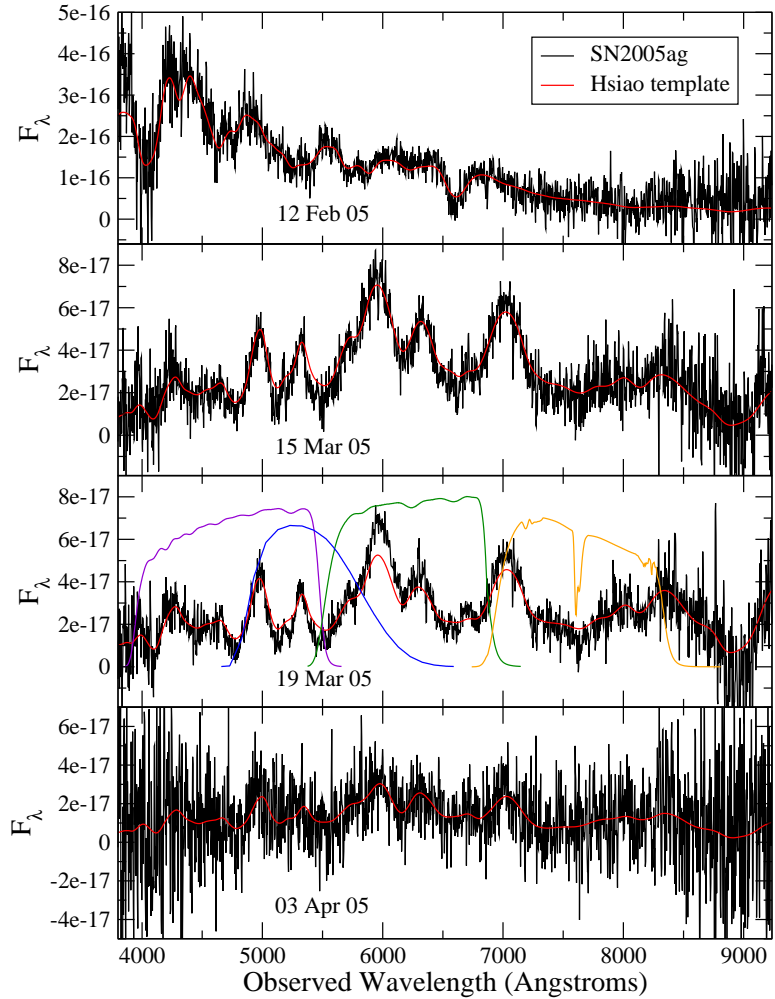}

\caption{Four spectra of SN~2005ag taking by the CSP (black lines) with the corresponding
SEDs from \citet{2007ApJ...663.1187H} (red lines).  The date of observation is labeled in
the panels.  The colored lines in the 3rd panel are the CSP filter functions:  $g$ (blue),
$r$ (yellow), $i$ (orange), and $V$ (green).\label{fig:spec_diff}}
\end{figure}

In order to compute a K-correction in filter $x$ for a SN~Ia with
heliocentric redshift $z_{hel}$, we use the following equation:
\begin{equation}
   K_{x}\left(t-t_{max}\left(B\right),z_{hel}\right)=2.5\log\left(1+z_{hel}\right)+
   2.5\log\left[\frac{\int F_{x}\left(\lambda\right)\Phi\left(\lambda;t-t_{max}\left(B\right)\right)\lambda d\lambda}
   {\int F_{x}\left(\lambda\right)\Phi\left(\lambda / \left(1+z_{hel}\right);t-t_{max}\left(B\right)\right)\lambda d\lambda}\right]
   \label{eq:k-correction}
\end{equation}
where $F_{x}\left(\lambda\right)$ is the response for filter $x$,
and $\Phi$ is the intrinsic SED of the SN~Ia. Unfortunately, due
to limited telescope resources, the SED of the SN~Ia is not observed
at every epoch, nor with sufficient wavelength coverage to compute
K-corrections for all filters. We therefore use the SED sequence developed
by \citet{2007ApJ...663.1187H} as a proxy. In order to account
for any possible reddening or intrinsic difference in the color of
the SN~Ia, we color-correct the SED by multiplying the original Hsiao
SED by a smooth function such that the observed colors match synthetic
colors derived from the corrected SED. \fig\ \ref{fig:K-corrections} shows
sample K-corrections derived from both the original Hsiao SED, as
well as one modified to match the observed colors, plotted as a function
of time for SN~2005ag. For the smooth function, we chose a tension
spline \citep{1987RenKa} with sufficient tension to produce the
observed colors to within approximately $\pm1\%$. The tension spline
gives us sufficient freedom to reproduce the observed colors, while
remaining more well-behaved than conventional splines. 
Also shown in \fig\ \ref{fig:K-corrections}
are the K-corrections derived from observed spectra of SN~2005ag obtained
by the CSP. In general, the agreement is excellent except for two epochs
in $V$ and $g$ band, near 30 days after $B$ maximum.  This is due to a difference
between the spectral features of the actual SED and the Hsiao SED.  \fig\ 
\ref{fig:spec_diff} shows the 4 spectra of SN~2005ag and the corresponding
Hsiao templates.  The observed spectrum shows a more prominent feature
near $\lambda = 6000$~\AA, which leads to the discrepancy in the K-correction.
This illustrates the limitations of using a single spectral template for
all SNe~Ia and the need for more spectra in order to construct a more
generalized SED sequence.

To compute the K-corrections, we therefore need to measure the time
of $B$-maximum, $t_{min}\left(B\right)$
and the observed colors $u-B$, $B-V$, $V-r$, $r-i$,
$i-Y$, $Y-J$, and $J-H$ as a function of time. To determine
$t_{min}\left(B\right)$, we fit the $B$-band light curve with a cubic
spline and solve for the time at which the derivative is zero. We
also fit splines to the other light curves in order to interpolate
any missing photometry when computing all the required colors.
This spline fitting procedure is further described in the following
section.

\subsection{Spline Fits}
\label{sec:splines}
Once the light curves have been K-corrected, we proceed to measure
$\dm\left(B\right)$ and the peak magnitudes in each filter.
For this, we fit cubic splines to the light curves, compute where
the derivative is zero, then use the spline to interpolate the brightness
of the light curve at that point. This allows us to measure $t_{max}\left(B\right)$,
which we use as the reference time for all the light curves. We can
then correct the light curves for time dilation by the factor $\left(1+z_{hel}\right)$.
Finally, we can interpolate the value of the $B$ light curve at 15
days after maximum in the frame of the SN~Ia, from which we compute
$\dm\left(B\right)$.

The business of fitting splines is a tricky one. Most spline algorithms
\citep[e.g.][]{1987RenKa,Dierckx1993} define some kind of smoothing
parameter (or tension) that allows the user to trade off between closeness
of fit and smoothness of the function. To remove the subjectiveness
of this, one can vary the smoothing parameter until the residuals
are consistent with the errors (e.g., $\chi_{\nu}\sim1$), but that
requires properly estimated errors (and covariances). Alternatively,
one can examine the statistical properties of the residuals. If the
spline is too smooth, then one expects that the residuals will be
correlated on some length-scale (i.e., several adjacent points will
be systematically under(over)-estimated by the spline, followed by several 
that are systematically over(under)-estimated by the spline).
Making the spline less and less smooth will decrease the auto-correlation
in the residuals.  \citet{297968} use the Durbin-Watson statistic
\citep{DurbinWatson}, which measures the degree of auto-correlation
in the residuals, to decide where to stop smoothing the spline
and we use their algorithm for interpolating the light curves.

Fitting splines is a non-linear process and in order to compute uncertainties
for the values of $t_{max}\left(B\right)$, $\dm\left(B\right)$,
and the peak magnitudes of the other filters, we perform Monte-Carlo simulations. 
The covariance
matrix of the photometry \citep[see ][]{2010AJ....139..519C} is used to
make realizations of the original light curves and each realization
is re-fit with splines and the light-curve parameters are re-computed. 
The second and fourth
columns of \tab\ \ref{tab:LCparams} list the decline rate parameter
$\Delta m_{15}(B)$ and $t_{min}\left(B\right)$
derived from the spline fits.  The errors are the
standard deviations of the Monte-Carlo-generated parameters.

\subsection{Light-Curve Templates and Errors}

\begin{figure}
\includegraphics[width=3in]{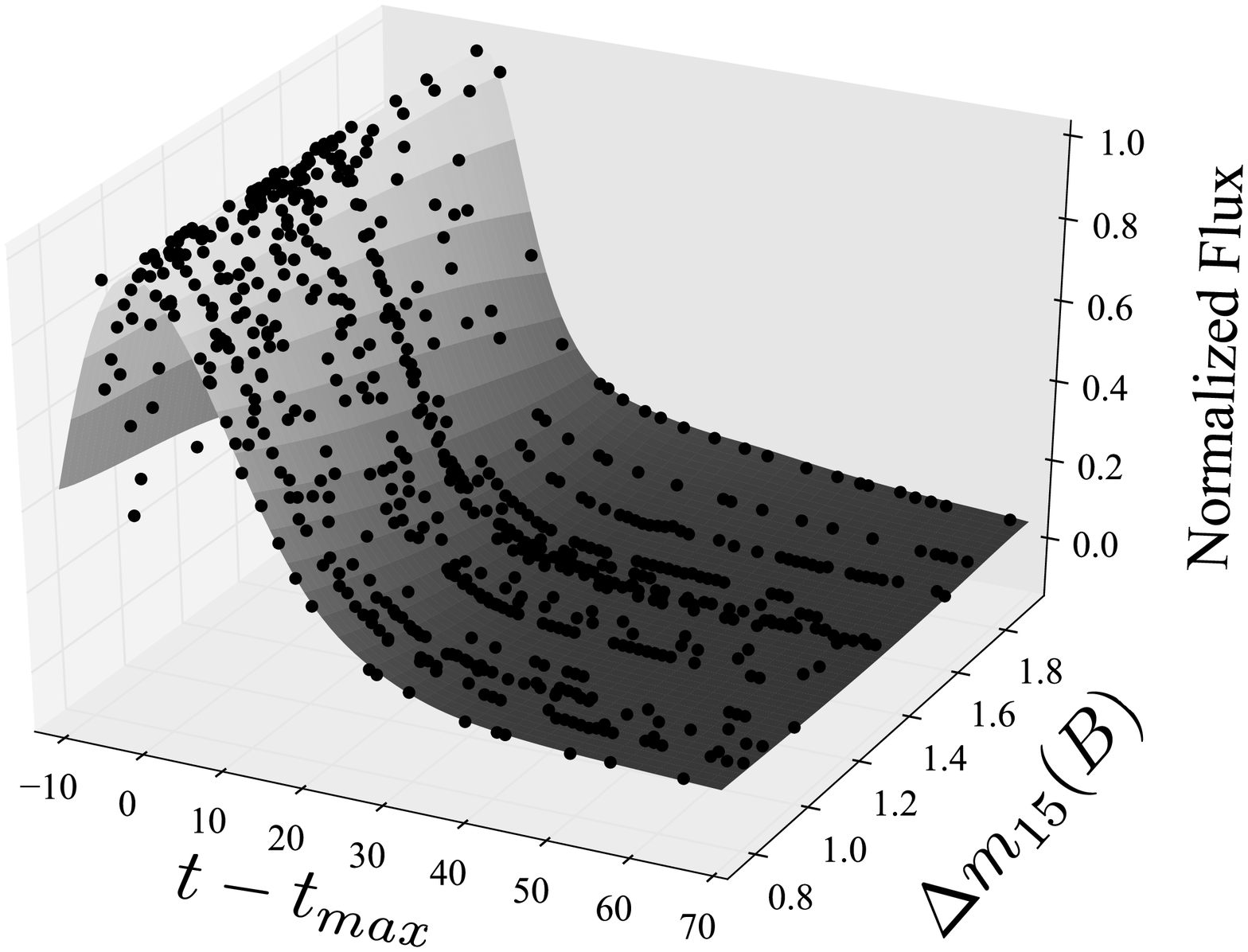}\includegraphics[width=3in]{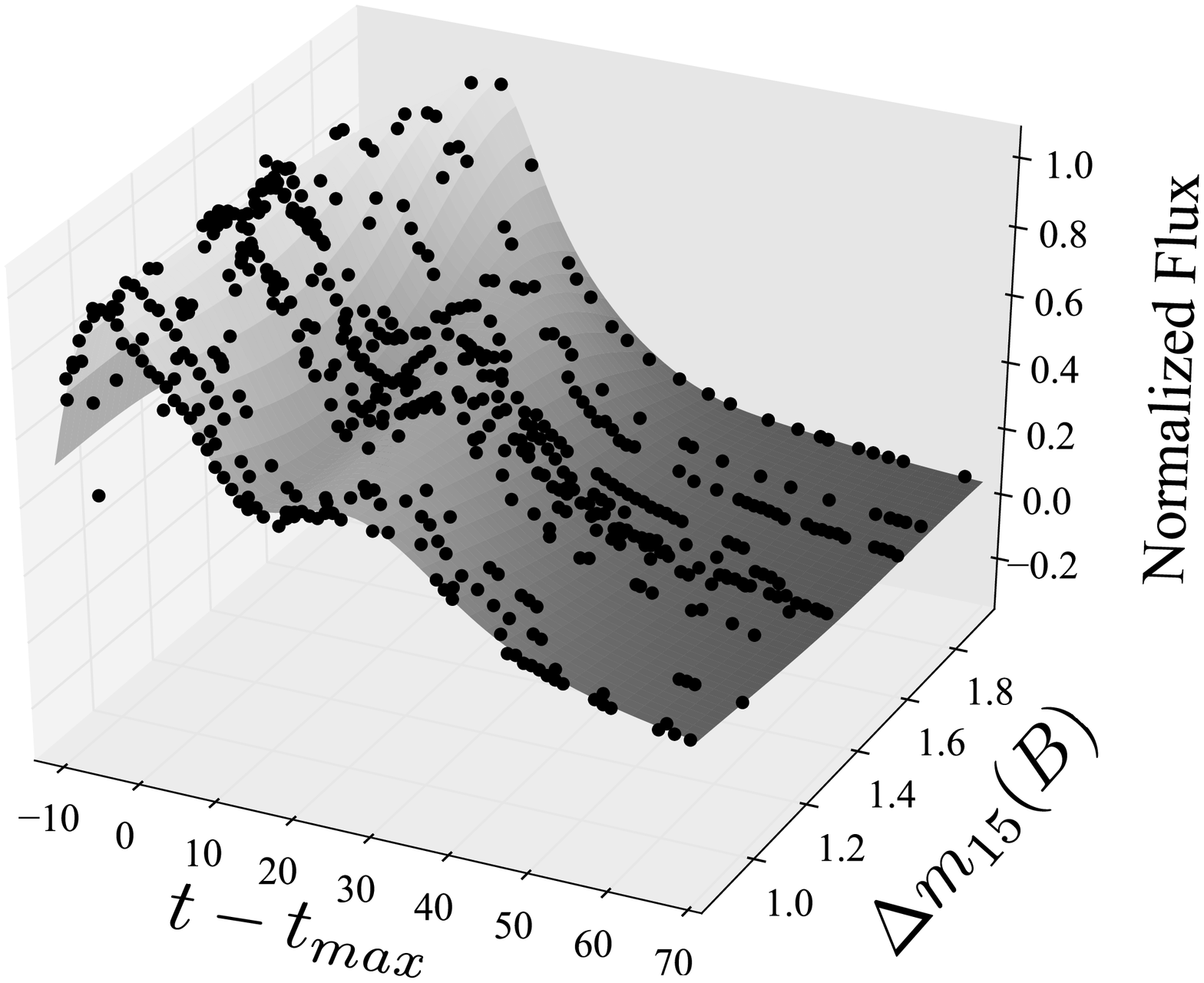}

\caption{The sparsely sampled 2D surface defined by the training set of SN
Ia. The three axes consist of (1) $t_{min}\left(B\right)$, (2)
the decline rate, $\dm$, and (3) the flux normalized to
the peak flux. Two filters are shown: $B$ band (left) and $i$ band (right).
\label{fig:template_surfaces} }

\end{figure}

We have a subset of \Ntemp\  local SNe~Ia  whose $B$-band light curve
has a well-defined peak observed in a set of filters $\{F_{j}\}$.
For each SN, we can measure $\dm(B)$ for each SN directly from the
$B$ light curve, giving us a set $\{\dm(B)_{i}\}$. Each
photometric data point then defines a coordinate 
$\left(\lambda_{j},t-t_{max,i}\left(B\right),\dm\left(B\right)_i,f_{j}/f_{j,max}\right)$,
where $\lambda_{j}$ is the effective wavelength of filter $j$, $t$
is the epoch of observation, $t_{max,i}\left(B\right)$ is the observed
time of maximum in the $B$ band%
\footnote{For simplicity, we have chosen $t_{min}\left(B\right)$ as reference
time for all filters. One could also measure independent reference
times for each filter and this will be investigated in the future.%
}, $f_j$ is the observed flux through filter $j$, and $f_{j,max}$ is the observed flux
at maximum. Together, these points define a sparsely-sampled 4-dimensional
surface. Generating a light-curve template can therefore be thought
of as interpolation on this surface. In fact, because filter band-passes
are more complicated than simple delta-functions, interpolation in
the $\lambda$ dimension is too simplistic a procedure and we must
resort to using S- and K-corrections when fitting obvserved data
at significant redshifts or in different passbands. Therefore, the problem reduces
to interpolation on a finite set of 3-dimensional surfaces, one for
each filter. \fig\ \ref{fig:template_surfaces} shows the distribution
of data for $B$ band and $i$ band for those SNe~Ia that have well-defined
peak magnitudes. 

Interpolation on a surface defined by sparsely-sampled data
(commonly referred to as \emph{Kriging}) is a common problem in science
and there exist many numerical solutions to the problem.
By inspection of the most densely sampled light curves and the seemingly gradual
evolution of the light-curve morphologies with $\Delta m_{15} (B)$, the underlying 
surface we wish to interpolate is very smooth and locally
can be approximated by a low-order polynomial. The GLOESS algorithm \citep{2004AJ....128.2239P},
a variant of the more well-known LOESS \citep{Cleveland1991} interpolator,
is therefore appropriate in this case and we use a 2-dimensional
extension which we call GLOESS2D.

\subsubsection{GLOESS2D}

GLOESS2D works by fitting a bi-variate polynomial function of order
$n$ to an observed set of $\left(x_{i},y_{i},z_{i}\right)$ points
and using this polynomial to interpolate $z_{0}$ at the point $\left(x_{0},y_{0}\right)$.
However, to make sure the interpolant reflects the underlying \emph{local}
trends of the data, the observed points are given the following weights:
\begin{equation}
   w_{i}=\frac{\exp\left[-\left(x_{i}-x_{0}\right)^{2}/2\sigma_{x}^{2}-
   \left(y_{i}-y_{0}\right)^{2}/2\sigma_{y}^{2}\right]}{\sigma_{i}^{2}},
\end{equation}
where $\sigma_{i}$ are the uncertainties in $z_{i}$, and $\sigma_{x}$
and $\sigma_{y}$ are the widths of the Gaussian window function.
In this way, $\sigma_{x}$ and $\sigma_{y}$ set the smoothing scales
in the x- and y-directions such that observed points for which $\left(x-x_{0}\right)\gg\sigma_{X}$
or $\left(y-y_{0}\right)\gg\sigma_{y}$ have very little effect on
the interpolation. 

The advantage of this approach over previous weighting schemes
\citep[e.g.][]{2006ApJ...647..501P} is that neighboring light curves
with similar $\dm\left(B\right)$
can serve to fill in temporal gaps in the observations. Nevertheless, 
the initial release
of the low-$z$ CSP data still has significant gaps in the light-curve
data, particularly at late times and for large values of $\dm\left(B\right)$.
This requires an adaptive weighting scheme. We construct a 2D metric
distance to each point on the surface: 
$d_{i}^{2}=a^{2}\left(t_{0}-t_{i}\right)^{2}+b^{2}\left(\Delta m_{15,0}-\Delta m_{15,i}\right)^{2}$
where $a$ and $b$ are arbitrary inverse scales. Each point on the
surface is then assigned a weight
\begin{equation}
   w_{i}=\frac{\exp\left(-d_{i}^{2}\right)}{\sigma_{i}^{2}},
\end{equation}
where $\sigma_{i}^{2}$ is the variance of the data point. This weight
is of course dependent on the scales $a$ and $b$, which in effect
control the ``smoothness'' of the interpolating function in either
dimension. Ideally, one would set these to some constant scale, however
our data at present have enough ``gaps'' that choosing too small
a scale leads to unstable fits, whereas choosing scales that are too
large results in templates that fail to capture the complex behavior
of the reddest filters. We have found that the data, at present, are
fit best with a constant $1/b=0.3$ and $1/a=3+0.1\left(t-t_{max}\right)$.
The latter captures the fact that observations are most densely sampled
in time near the peak and that the light curves follow an exponential
decay at late times. As our sample grows and the gaps fill in, this
functional form for $a$ will no longer be necessary.

Using this interpolating algorithm, we can fit a smooth surface for
each filter. Generating light-curve templates can then be done by
sampling along a constant-$\dm(B)$ slice of these surfaces
at arbitrary resolution. More useful for SNooPy is simply interpolating
the epochs at which the SN~Ia was observed for a given value of $\dm$
as part of a least-squares fitting routine.

\begin{figure}
\includegraphics[width=3in]{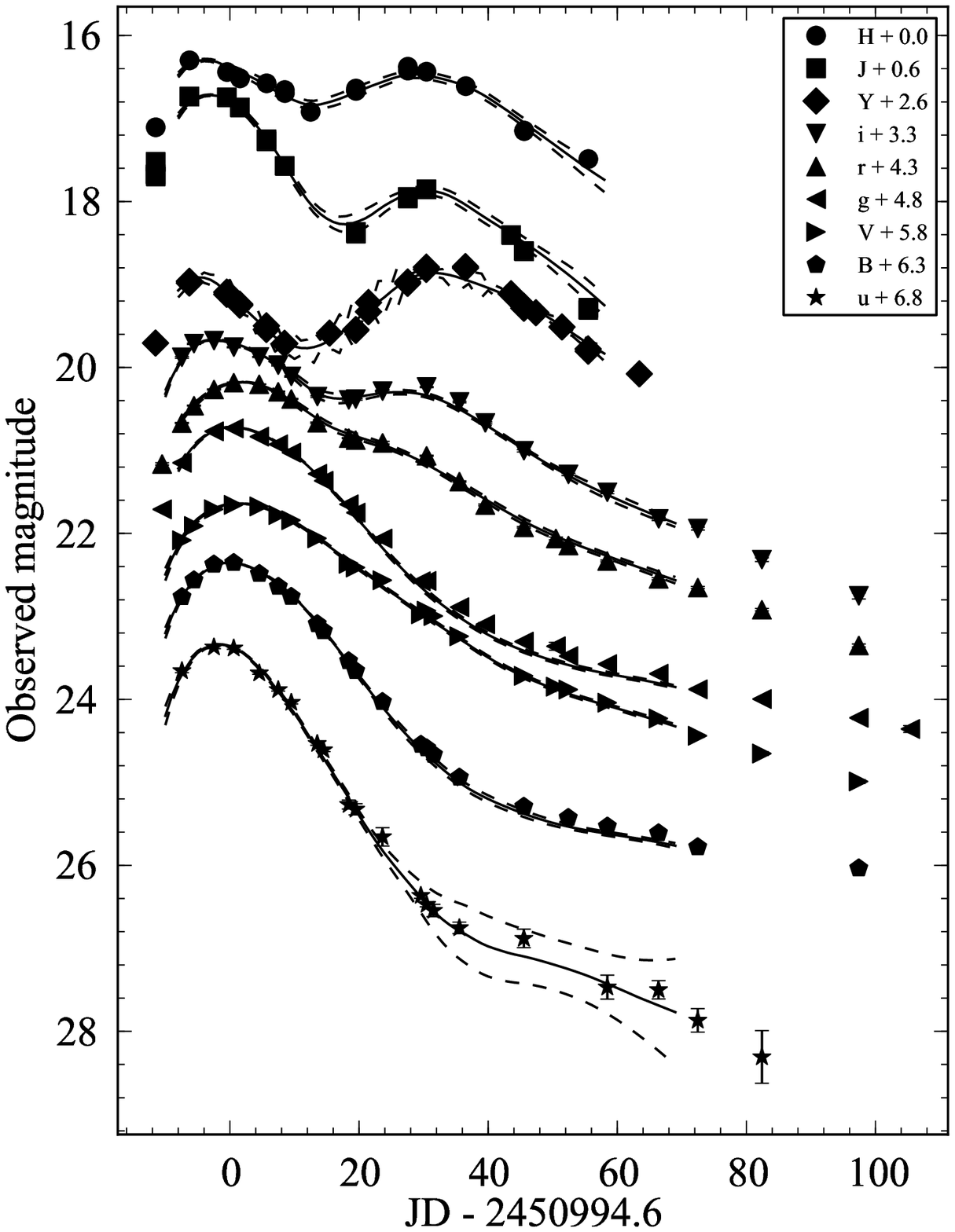}\includegraphics[width=3in]{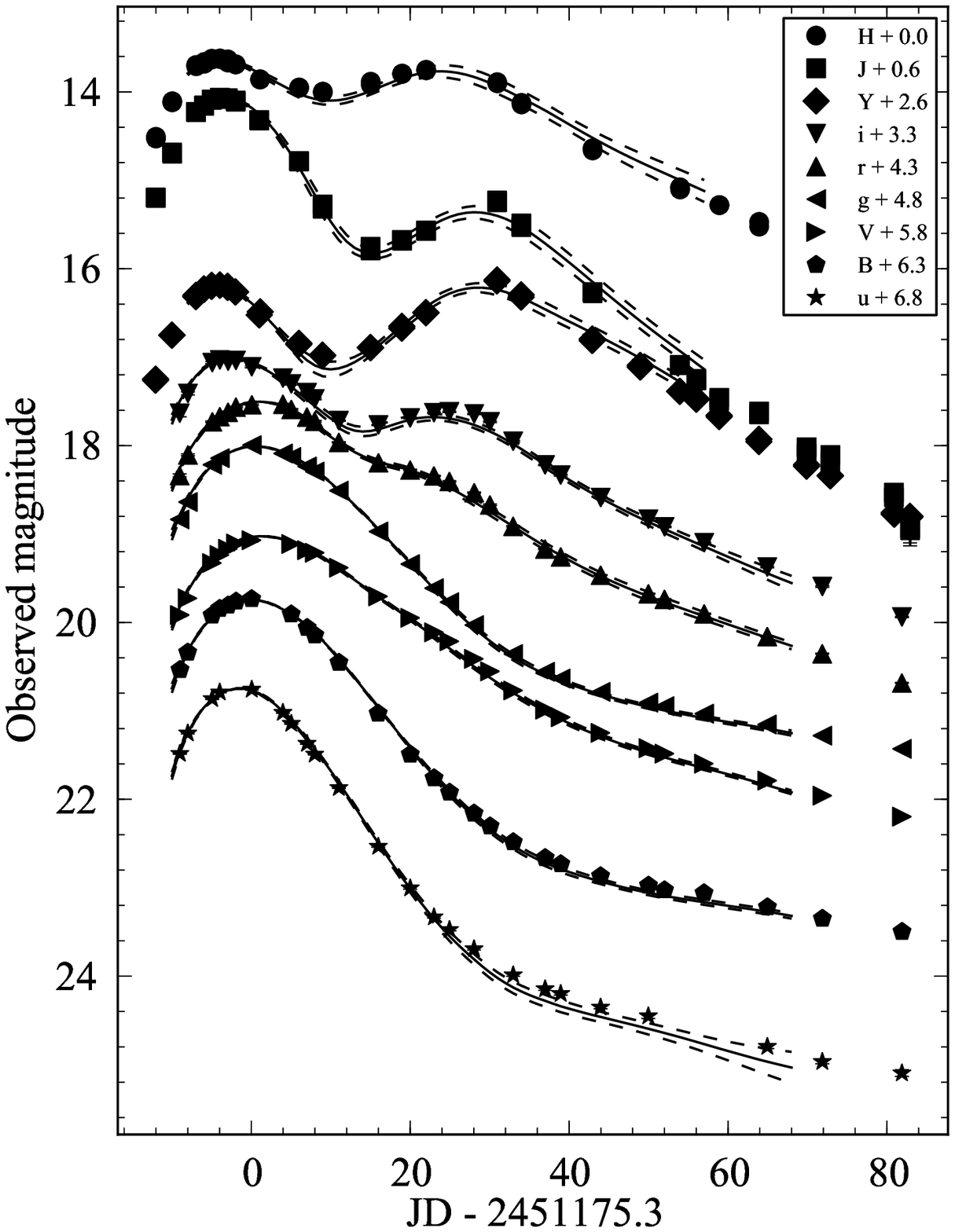}
\caption{Sample light curve fits using SNooPy for two SN~Ia:  SN~2006et (left) and 
SN~2007af (right).
The data points are offset by the amount indicated in the legends for clarity.  The solid lines
are fits to the photometry, the dashed lines are the 1-$\sigma$ template errors.  \label{fig:sample-lc}}
\end{figure}

Note that there is no ``training'' involved in this procedure.
We are simply interpolating on a surface defined by a pre-existing
set of SNe~Ia. Data can be added or subtracted at will and the effects
are immediate (a fact we exploit in the next section). 
Sample light-curve fits using SNooPy templates can be found in
\citet{2010AJ....139..519C}.  In \fig\ \ref{fig:sample-lc}, we show light-curve
fits for the two SNe~Ia added to the sample for this paper:  SN~2006et and
SN~2007af.

\subsubsection{The Meaning of \dm \label{sub:dm15_defined}}
Now that we have the ability to create light-curve templates in any of the 
CSP filter bandpasses, the meaning of the decline rate parameter is not
so clear.  Originally, we defined $\Delta m_{15}(B)$ as the change in
magnitude of the $B$-band between maximum and 15 days after maximum.  If we use the
above procedure to generate a $B$ light-curve template with a particular
$\dm(B)$, there is no guarantee that the {\em measured} change in magnitude
of the template from peak to day 15 will exactly match the input $\dm(B)$.  Furthermore, when
fitting light curves with templates, all the data from all the filters contribute to
the solution, not just $B$-band data close to maximum and day 15.  For this
reason, the template-derived value of the decline rate parameter, which we denote
as simply $\dm$, can deviate both randomly and systematically from the directly
measured value, $\dm(B)$.

\begin{figure}
\includegraphics[width=6in]{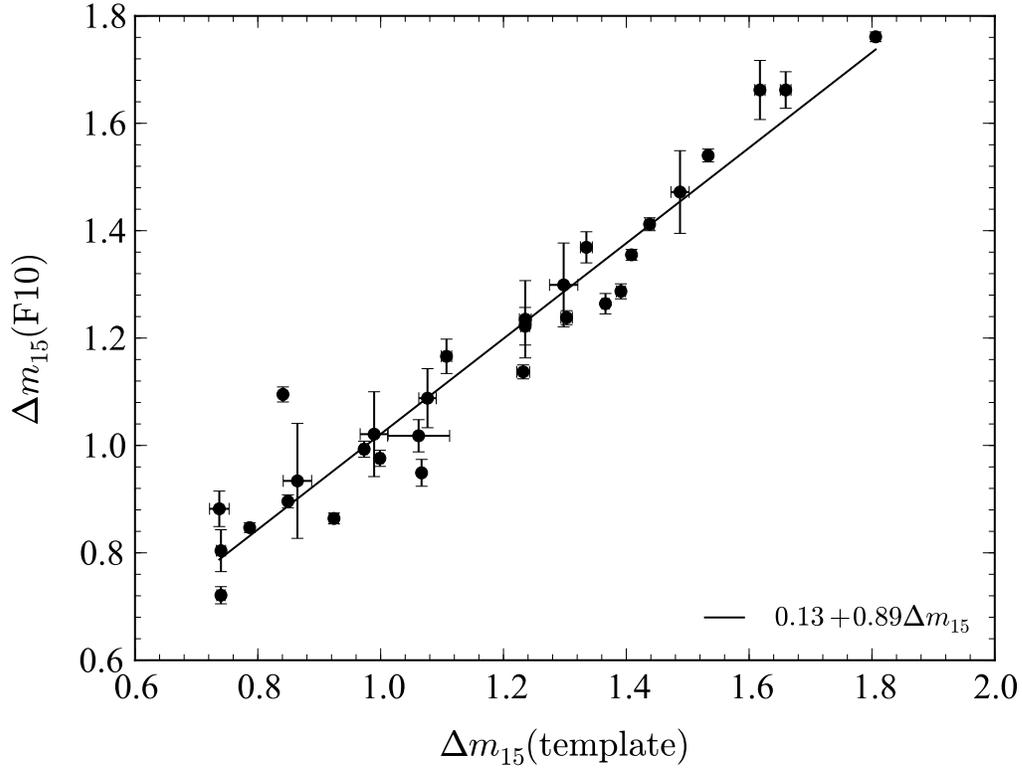}
\caption{Comparison of the decline rate parameter as measured by F10, $\dm(B)$ with the value determined
through template fitting.  The solid line shows the best linear fit to the data.\label{fig:compare_dm15s}}
\end{figure}

\fig\ \ref{fig:compare_dm15s} shows a comparison between the value of $\dm$
derived from template fits to those determined by F10.  While there is random
scatter as expected, there is also a systematic trend as shown by the solid
line, which has less than unit slope.  This systematic difference between
$\dm(B)$ and $\dm$ will have an impact on the calibration of our SN~Ia sample
for determining distances.  More precisely,  we can expect that the $\dm$
luminosity correction will have a smaller coefficient than the $\dm(B)$
luminosity correction.  
This is what we find in \S \ref{sub:Tripp-Model} and \ref{sub:EBV_model}.

\subsubsection{Template Errors\label{sub:Template-Errors}}

\begin{figure}
\includegraphics[width=3in]{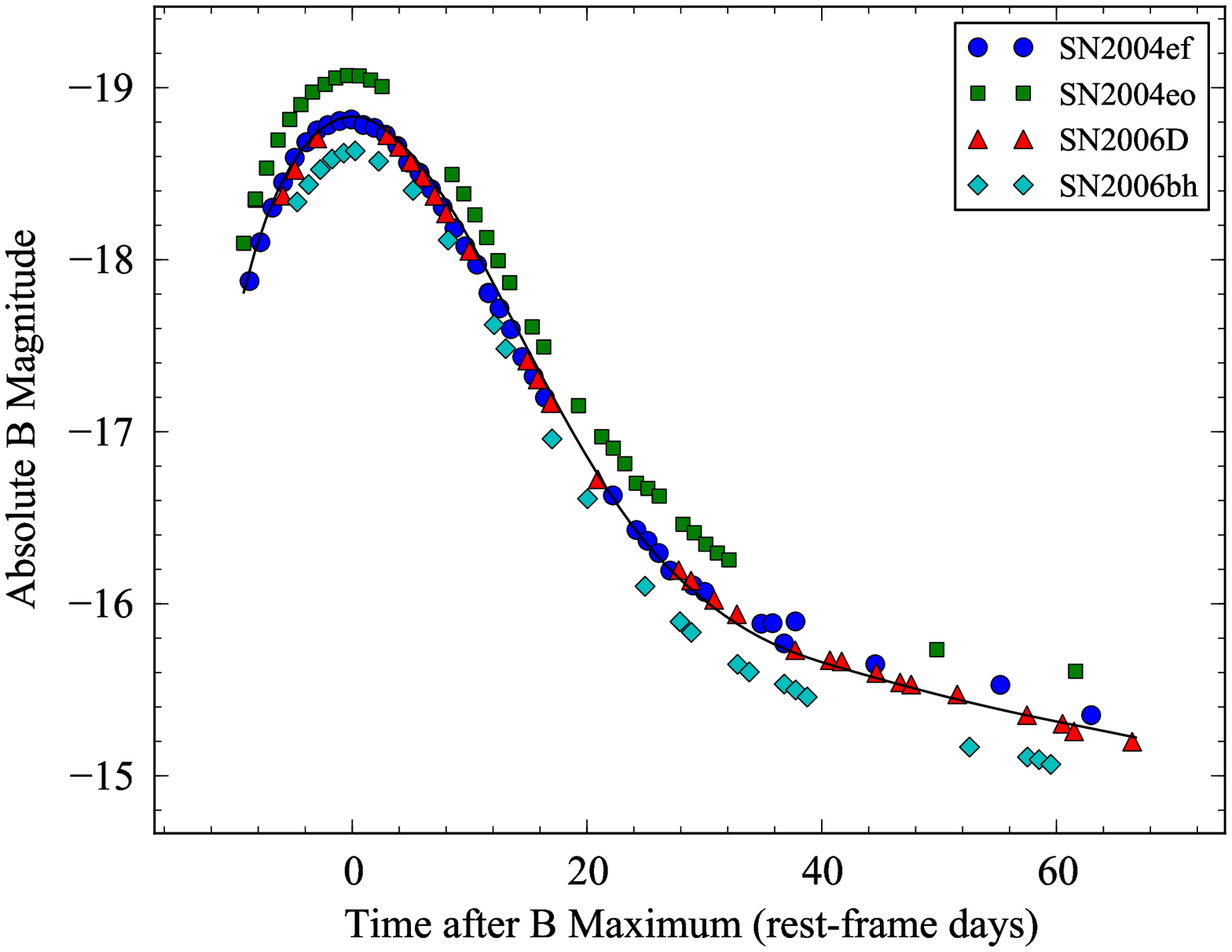}\includegraphics[width=3in]{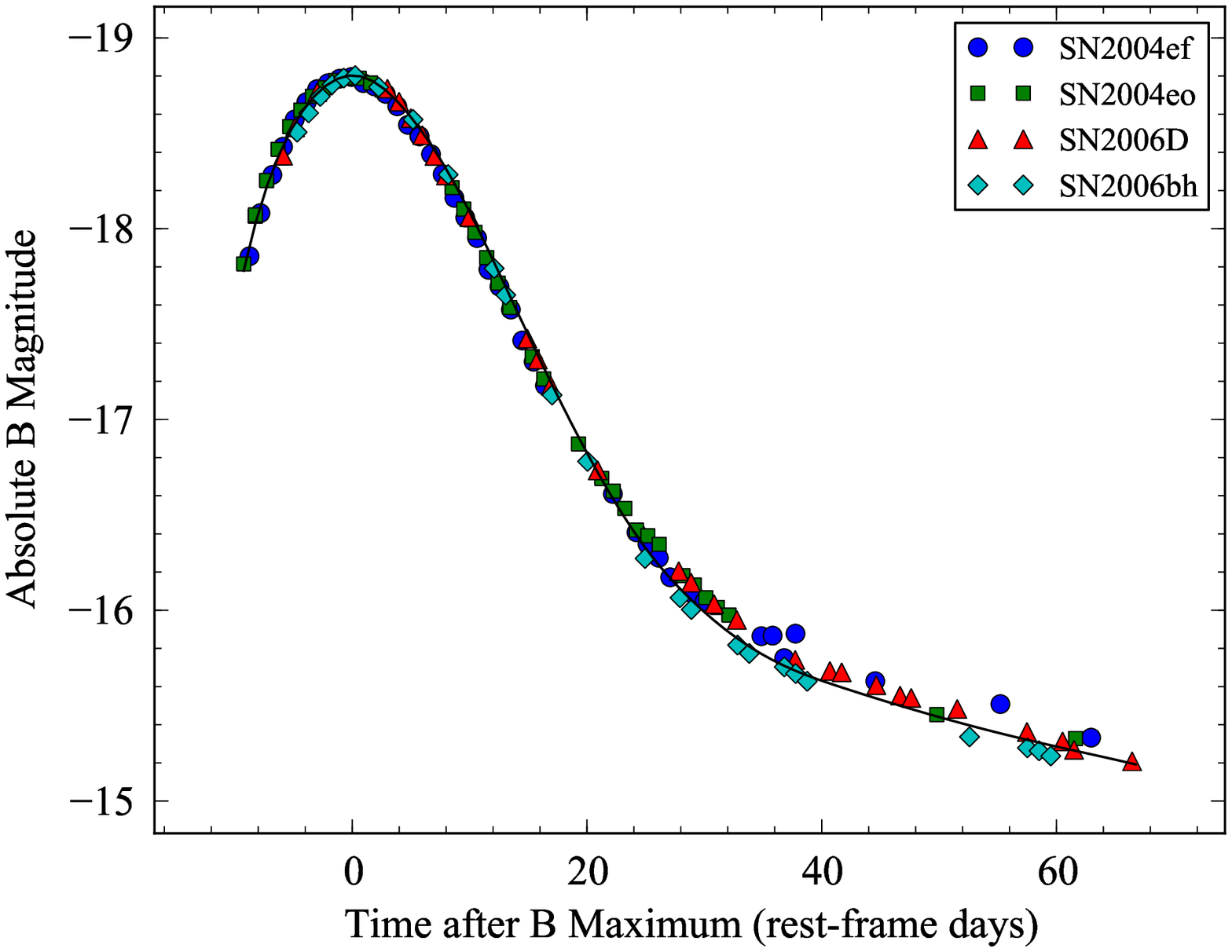}

\caption{Absolute magnitude $B$-band light curves for four SNe~Ia with
similar \dm: SN~2004ef ($\dm = 1.36$),
SN~2004eo ($\dm = 1.38$), SN~2006D ($\dm = 1.38$), and SN~2006bh ($\dm = 1.43$). 
Left: the distances are computed using
a Virgo-corrected velocity and Hubble constant of $72\ {\rm km}\cdot {\rm s}^{-1}\cdot {\rm Mpc}^{-1}$.
Right:  the light curves are shifted such that the $B$-maxima agree.  The solid
lines are SNooPy $B$-band templates for $\dm = 1.4$.\label{fig:comp-lcs-B}}

\end{figure}

\begin{figure}
\includegraphics[width=3in]{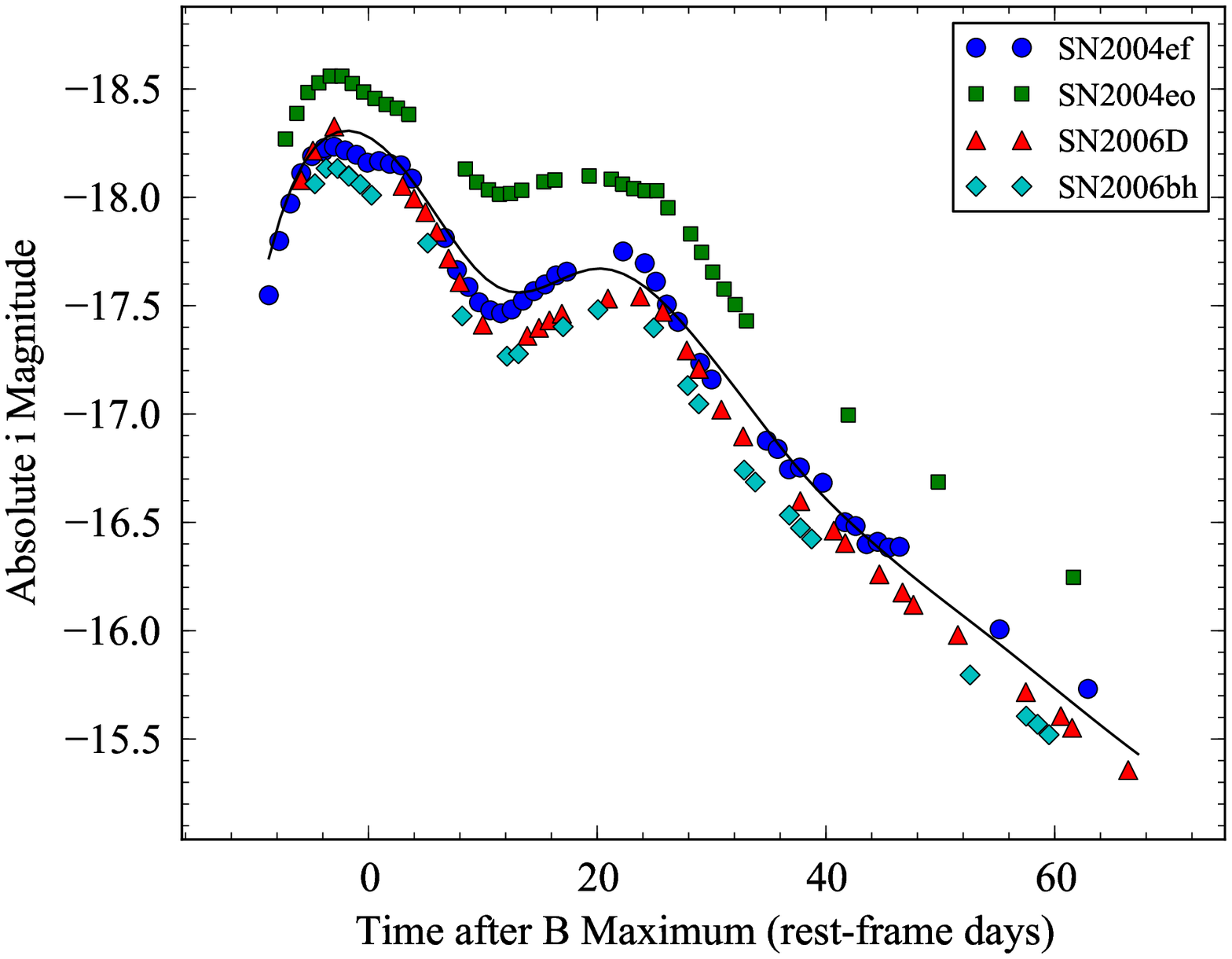}\includegraphics[width=3in]{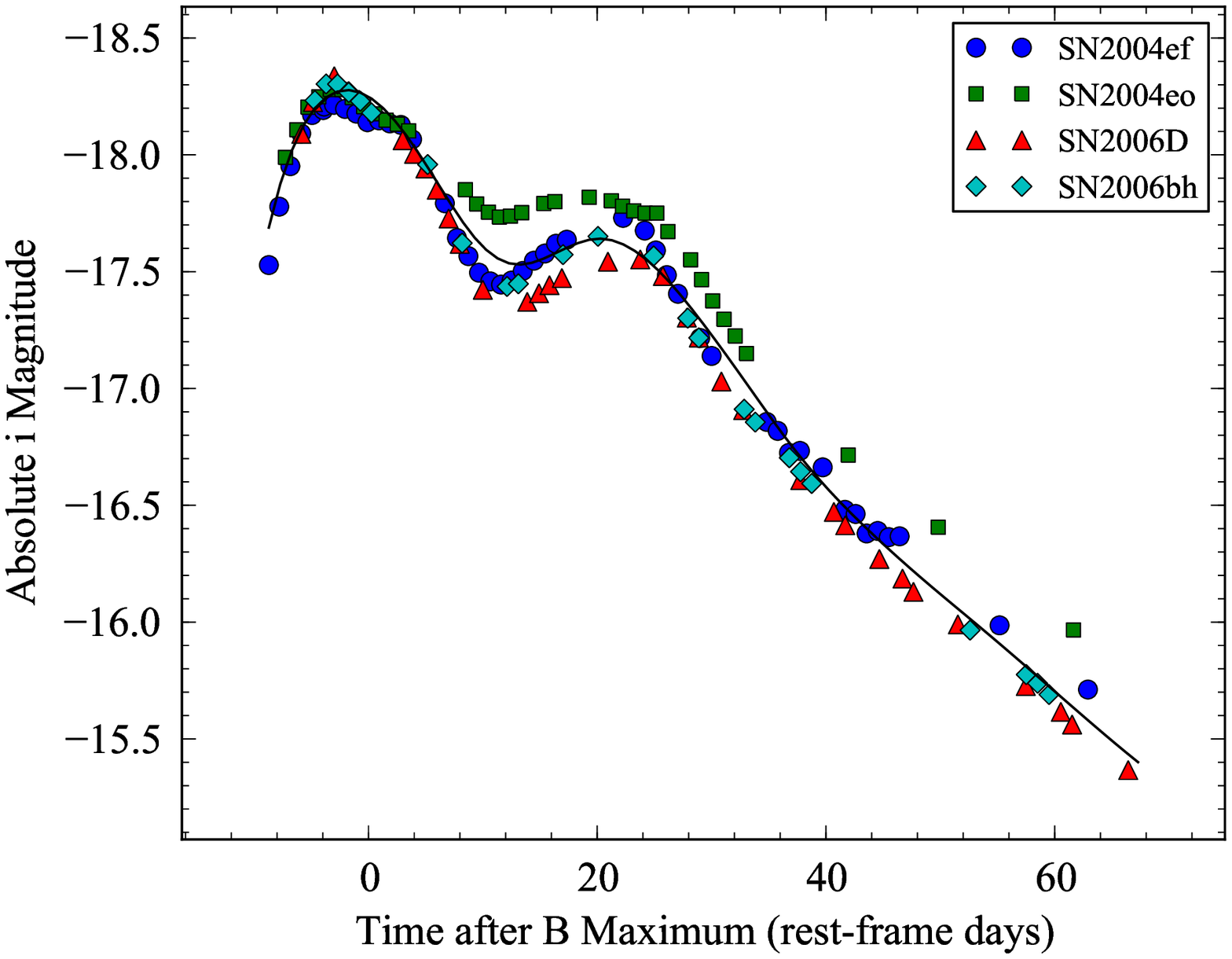}

\caption{Absolute magnitude $i$-band light curves for the same SNe as in \fig\
\ref{fig:comp-lcs-B}. The distances in the left panel and offsets in right
panel are
the same as those used in \fig\ \ref{fig:comp-lcs-B}.  The solid lines are
SNooPy $i$-band templates for $\dm = 1.4$.\label{fig:comp-lcs-i}}

\end{figure}

Aside from exhibiting a more complicated light-curve morphology, it
has been empirically determined that SNe~Ia show an marked increase
in the variation in the light-curve behavior at longer wavelengths
for a given decline rate. In particular, \citet{2010AJ....139..120F} displayed
the varied behavior of 4 SNe~Ia with very similar $\dm\left(B\right)$:
SN~2004eo, SN~2004ef, SN~2006D, and SN~2006bh. In their \fig\ 8, the
light curves in $B$, $V$, and $i$ were compared by normalizing to the maximum
of each light curve. An important question immediately arises: is
the time of maximum truly the epoch at which the dispersion is minimized?
In order to address this, we reproduce F10's \fig\ 8, but instead
of normalizing to any particular epoch, we simply plot absolute magnitudes
rather than apparent. To do this, we use
distance moduli based on a Virgo-corrected redshift obtained from
the NASA/IPAC Extragalactic Database (NED), assuming a Hubble constant
$H_{o}=72\ {\rm km\cdot s^{-1}\cdot Mpc^{-1}}$, and by correcting for reddening
assuming values of $E\left(B-V\right)$ from F10. These absolute
magnitude
light curves are plotted in left left panels of \fig s \ref{fig:comp-lcs-B}
and \ref{fig:comp-lcs-i}. Due to errors in distance, peculiar velocities,
and/or differences
in intrinsic luminosity, the light curves do not match up exactly,
however it can be seen that the differences are correlated between
$B$  and $i$ bands. It is also intriguing that SN~2004eo, whose $i$ band
light curve is significantly shallower than the other 3, is more luminous by about
0.3 mag.  If we compute offsets such that the $B$-band light curves overlap (i.e.,
insisting that the 4 SNe are standard candles in $B$) and apply these
same offsets to the $i$ band, it is readily seen that the i-band
light curves do indeed have their best agreement near the time of
maximum. This is shown in the right panels of \fig s \ref{fig:comp-lcs-B}
and \ref{fig:comp-lcs-i}. We also over-plot the SNooPy template for
the average $\dm$ for this sample. The template light curve,
as expected, traces the average behavior of these 4 SNe~Ia.  Unfortunately,
the CSP only observed one (SN~2006bh) spectroscopically in the time interval
where the 4 SNe~Ia are most discrepant, so we cannot determine
directly what causes these differences.  However, $i$-band
covers a prominent spectra feature:  the \ion{Ca}{2} triplet, which is known
to vary greatly from SN to SN, even those with similar \dm.
These variations are somewhat related to the velocity structure of
the ejecta and will be presented in greater detail in an upcoming
CSP paper (Folatelli, in prep.)

On the face of it, the $i$ and NIR light curves present a problem
for template-fitting: a single parameter family of curves seems inadequate
for properly reproducing the observations away from maximum light.
On the other hand, this opens up the possibility that there is at
least one more parameter one could employ to improve the fits, which
could potentially increase our knowledge of these events or even make
them better standard candles. 

It is beyond the scope of this paper (and this relatively small data
set) to solve for this second light-curve parameter, but this will be 
investigated
further as the CSP sample increases. For the time being, however,
we need to incorporate these intrinsic variations as extra error in
the template. The reason for this is to both ensure that the final
statistical errors in the light-curve parameters reflect these intrinsic
variations and also to weight the data near maximum more than at later
times. 

To estimate the extra dispersion, we use a bootstrap technique.
From the \Ntemp\ SNe~Ia that were used to construct the templates, we randomly
choose the same number, with replacement, generating 100 different
sample realizations. The templates are then re-generated from these
realizations and residuals from the master are computed. We then take
the rms scatter about the master as the 1-$\sigma$ error in the template.

\begin{figure}
\includegraphics[width=5in]{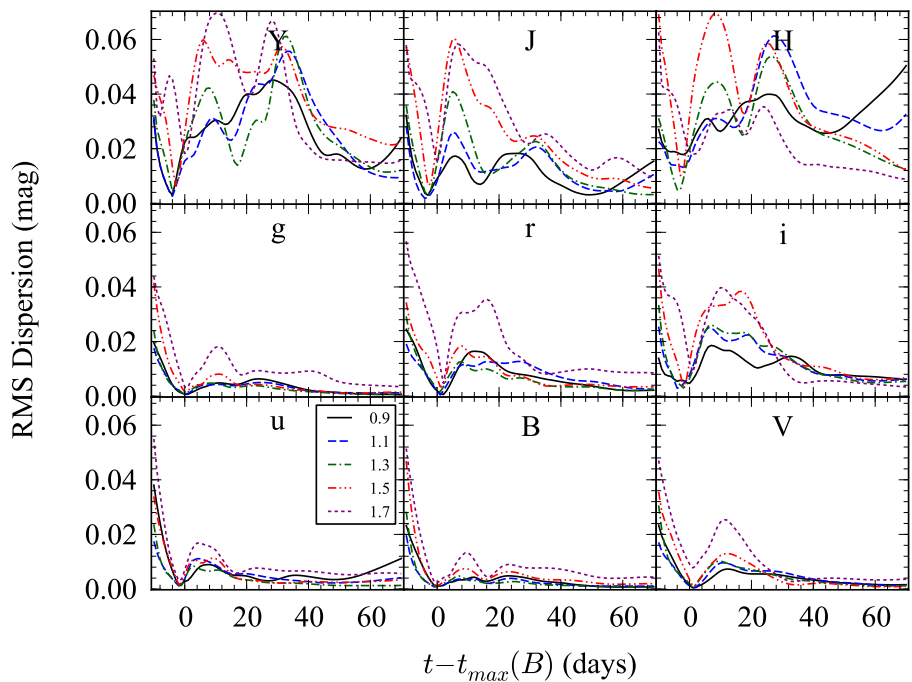}

\caption{Results of a bootstrap simulation on the dispersion in the light-curve
templates due to intrinsic variations from supernova to supernova. Each panel represents
a separate CSP filter and the different lines represent different
values of $\dm$. Each line shows the rms dispersion of
the templates as a function of time since $B$-maximum.\label{fig:BS-RMS}}

\end{figure}

\begin{figure}
\includegraphics[width=5in]{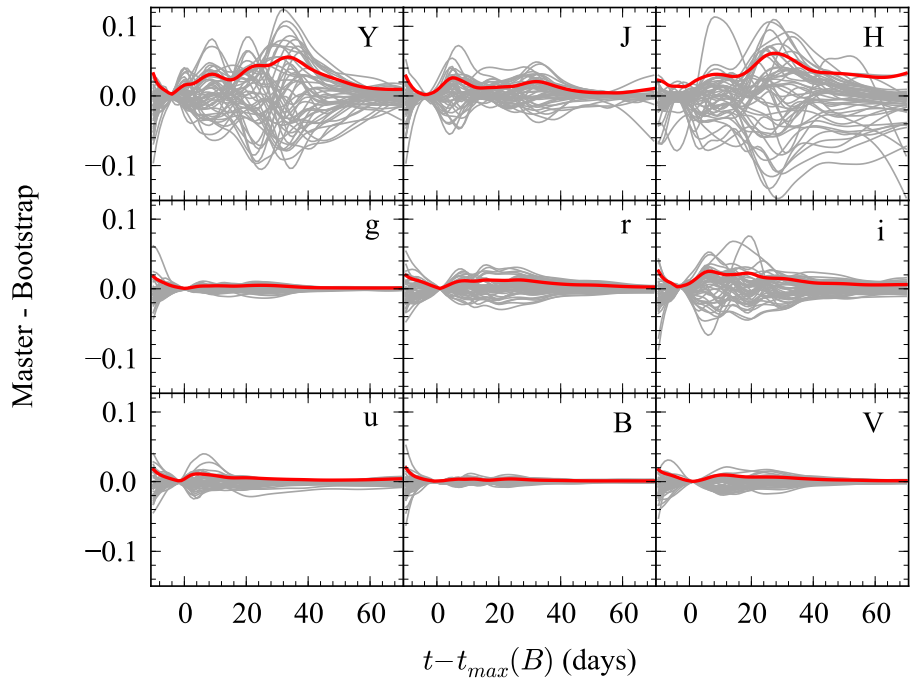}

\caption{Same as \fig\ \ref{fig:BS-RMS}, but for a specific value of $\dm=1.1$.
The grey lines represent the individual bootstrap realizations. The
red line represents the rms of all the realizations.\label{fig:BS-1.1}}

\end{figure}

The results of this bootstrap method are shown in \fig\ \ref{fig:BS-RMS},
where we plot the 1-$\sigma$ errors as a function of time and $\dm$.
In \fig\ \ref{fig:BS-1.1}, we show the special case of $\dm=1.1$
in more detail, plotting each realization as a gray line and the rms
line in red.  We also plot the template errors as dashed lines in \fig\
\ref{fig:sample-lc}.  It is immediately apparent that the $i$ and NIR light curves
show greater dispersion than the optical.  At least part of the reason for
this is the smaller number of SNe~Ia for which we have sufficient NIR coverage,
particularly at the extreme ends of the \dm\ distribution.
For example, of the \Ntemp\ SNe~Ia that are used to generate templates, only 
\Ntempi\ were
suitable for constructing the $i$-band templates and only \NtempNIR\ could be 
used for the NIR templates.  Indeed, this lack of suitable light curves
is what lead us to use SN~2006X for the NIR templates, despite the fact
it is known to have a light-echo in the bluer bands.  So although it is
clear the NIR suffers from
more intrinsic variation in light-curve morphology than the optical 
wavelengths, this method currently overestimates it.  The same problem
occurs for all filters at the high-\dm\ end, where the small number of
SNe~Ia artificially increases the errors relative to other
values of \dm.  The addition of more SNe from the later CSP campaigns
will greatly improve the estimates of these errors.

\subsubsection{Extrapolation Errors}

Due to the fact that the intrinsic variations in the templates 
seem to increase
at later times, using them to fit data which begin after the maximum
occurred will likely introduce additional error when extrapolating
back to the maximum. The magnitude of these errors will of course
depend on the filter and on how late the observations start. We therefore
estimate the extrapolation errors in the following manner: 1) for
each filter, we assemble those light curves from the sample
whose peak magnitudes are well established and whose light curve is
sampled to at least 30 days after $B$-maximum; 2) we eliminate any
data earlier than $t_{cut}$ days after the time of $B$-maximum;
3) we fit this new data using templates and compare the peak magnitudes,
$t_{min}\left(B\right)$ and \dm\ to the originals. 

\begin{figure}
\includegraphics[width=6in]{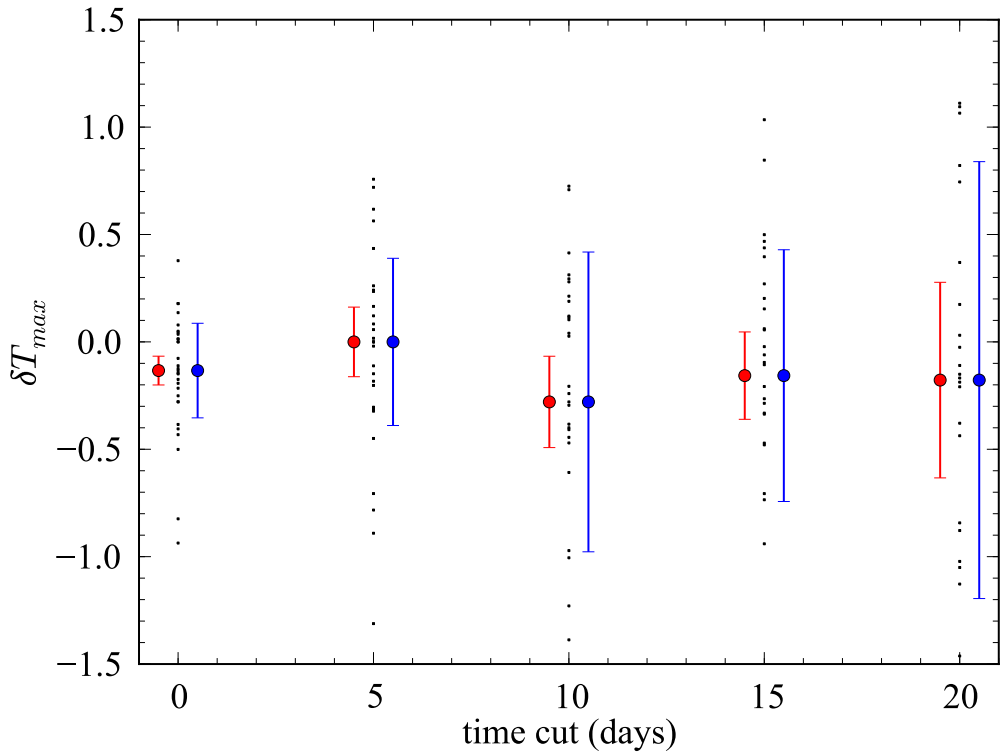}

\caption{Errors in the value of $t_{max}$ as a function of the time of earliest
observation ($t_{cut}$). Each SN~Ia is plotted as a black dot. The blue
circles and error bars to the right of each set of dots correspond to the 
median and median absolute
deviation, respectively. The red circles and error bars to the left of
each set shows the weighed mean
deviation after the least-squares error in $t_{max}$ has been removed.
\label{fig:tmax-extrap}}

\end{figure}

\begin{figure}
\includegraphics[width=6in]{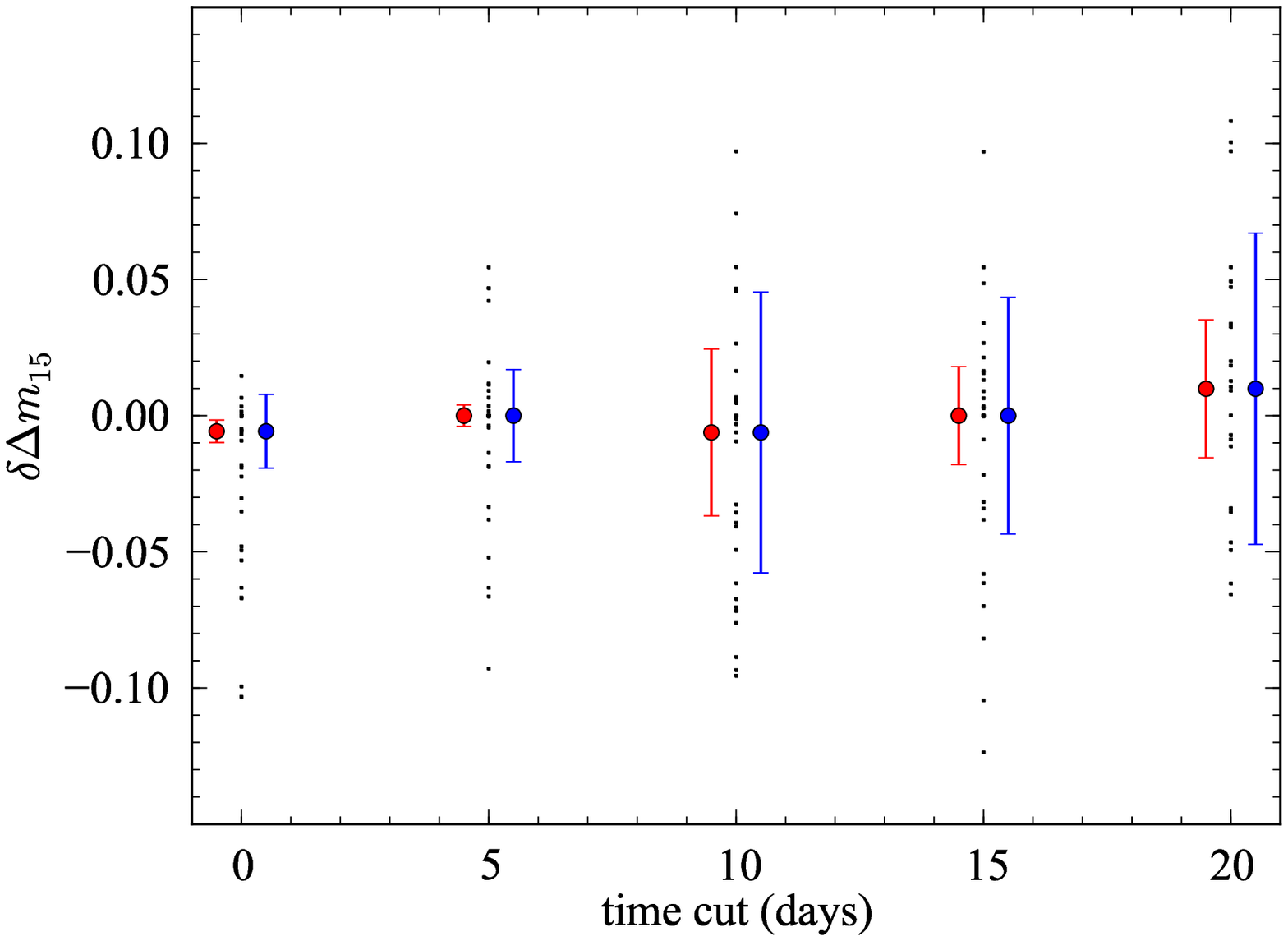}

\caption{Same as \fig\ \ref{fig:tmax-extrap}, but for errors in $\dm$.\label{fig:dm15-extrap}}

\end{figure}

\begin{figure}
\includegraphics[width=6in]{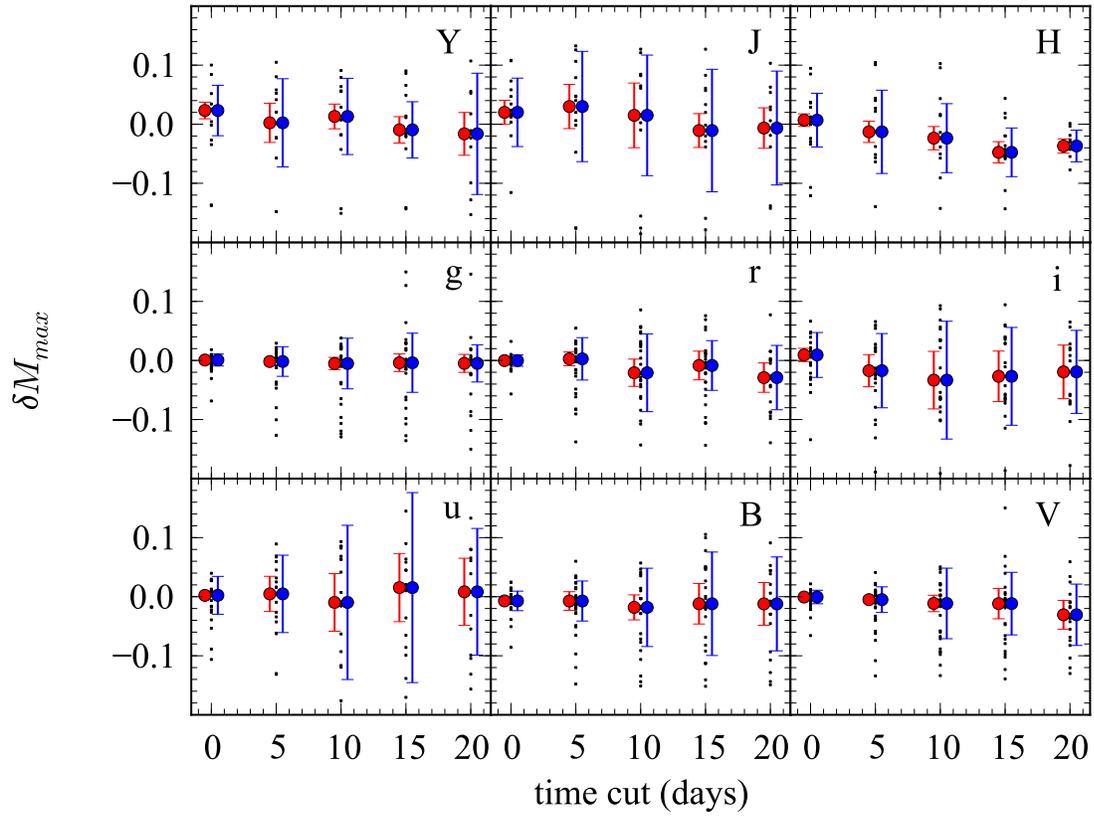}

\caption{Same as \fig\ \ref{fig:tmax-extrap}, but for errors in apparent
maximum magnitude. Each panel represents a different filter. \label{fig:mmax-extrap}}

\end{figure}

The results of these tests are shown in \fig s \ref{fig:tmax-extrap},
\ref{fig:dm15-extrap}, and
\ref{fig:mmax-extrap}. In each figure, we plot the differences
between the fit parameter using the cut light curves and the original
parameter value as a function of cut time $t_{cut}$. Each individual
SN is plotted as a black point. We then plot the median as a blue
point and the median absolute deviation as the blue error-bar. However,
to measure the \emph{excess} error due to extrapolation, we also compute
the median of the absolute deviation minus the error reported by the
least-squares fit. These errors are plotted as red error-bars. In
almost all cases, the median residuals are consistent with zero, indicating
no systematic error in the extrapolated magnitudes at maximum, nor
any systematic error in the determination of $\dm$ or $t_{max}$.
However, we do find significant excess error due to the extrapolation
as $t_{cut}$ becomes larger. In general, the errors seem to grow
rapidly until 10 days after $B$-maximum, then level off or even
decrease.  This is due to the increased error in the templates
near day 10 (see \fig\ \ref{fig:BS-RMS}). \tab\ \ref{tab:extrap-error}
lists the excess errors that are adopted for use in SNooPy.

\section{Distance Estimations}
\label{sec:distances}
With template light curves in hand, we turn to the task of using them
to fit a distance to a SN~Ia that is not in the calibrating sample.
To do this, we propose the following model for the observed magnitude
of a SN~Ia observed in filter $x$:
\begin{equation}
m_{x}\left(t\right)=T_{x}\left(t-t_{max};\dm\right)+
M_{x}\left(\dm\right)+f\left(y-z\right)+\mu\label{eq:DM},
\end{equation}
where $T_{x}$ is the light-curve template, parameterized by $\dm$,
$M_{x}$ is the absolute magnitude of a SN~Ia in filter $x$, $f\left(y-z\right)$
is some function of the SN~Ia's color defined by filters $y$ and $z$,
and $\mu$ is the desired distance modulus. Using the results of F10, 
we assume a linear relationship
$M_{x}\left(\dm\right)=a_{x}+b_{x}\left(\dm-1.1\right)$
and a sample of low-$z$ SN~Ia are used to determine $a_{x}$
and $b_{x}$ for each filter $x$. The choice of $f\left(y-z\right)$
usually consists of either an empirical correction proportional to
the color $y-z$ or a correction for an assumed extinction $E\left(y-z\right)$
due to dust. Both have advantages and disadvantages and we examine
each in turn.

\subsection{Reddening as a Parameter\label{sub:red_param}}

The distribution of observed colors of SNe~Ia are usually attributed 
to two causes:
an intrinsic color that is correlated with the width of the light curve
(the so-called redder-dimmer relation) and the extinction by dust
in both the Milky Way and the host galaxy (as well as any possible dust
in the IGM or surrounding the supernova itself). Unfortunately, these
two effects work roughly in parallel and it is difficult to disentangle
what is reddening due to dust and what is an intrinsically red SN~Ia.
To separate the effects, one must isolate a subsample of objects
that are believed to be ``unreddened'' by dust, and analyze the
intrinsic color distribution of this subsample. This was done by F10,
who derived simple linear formulae for the intrinsic colors of SNe~Ia:
$\left(y-z\right)_{0}=a_{yz}+b_{yz}\left(\dm-1.1\right)$.
Using these intrinsic colors, they could then compute $E\left(y-z\right)$
color excesses and hence extinction corrections for their entire sample.
For each band $x$, they then solved for the absolute magnitude as
a function of decline rate: \begin{equation}
M_{x}=a_{x}+b_{x}\left(\dm-1.1\right)\label{eq:AbsMag}\end{equation}
which encapsulates both the brightness-$\dm$ relation as
well as the color-$\dm$ relation. With this calibration,
we can construct a model of the observed light curve of any other SN~Ia:
\begin{equation}
m_{x}\left(t\right)=T_{x}\left(t-t_{max};\dm\right)+a_{x}+
b_{x}\left(\dm-1.1\right)+R_{x}^{BV}E\left(B-V\right)+
\mu\label{eq:DM_EBV}
\end{equation}
where $R_{x}^{BV}$ is the ratio of total to selective absorption,
defined as
\begin{equation}
R_{x}^{yz}\equiv\frac{A_{x}}{A_{y}-A_{z}}=\frac{A_{x}}{E(y-z)}\label{eq:red_law}.
\end{equation}
Naturally, at least two different bands must be observed for each
SN~Ia, otherwise $E\left(B-V\right)$ is degenerate with $\mu$.

Because SNe~Ia have SEDs that are significantly non-stellar, the 
reddening coefficient
will not follow standard extinction curves (e.g., CCM). Instead, the
reddening coefficient $R_{x}^{yz}$ can be computed by 1) assuming
an appropriately red-shifted SED for the SN~Ia, 2) applying an extinction
curve $A\left(\lambda\right)$ to the SED, 3) using the known filter
functions to compute synthetic extinctions in each filter, and 4)
using equation (\ref{eq:red_law}). We use the extinction curve derived
by CCM+O, which can be parameterized by the reddening coefficient in the
Johnson $V$ band ($R_V$), and the $E\left(B-V\right)$ color excess. Several
recent studies have shown that SNe~Ia seem to ``prefer'' a low
value of $R_{V}$ compared to the Milky Way average
\citep{1999ApJ...525..209T,2006A&A...447...31A,2009ApJ...700.1097H}. 
This could indicate
that SNe~Ia reside in peculiar environments, or that there remains
a luminosity-color relation that is independent of $\dm$.
F10 have tabulated several different values of $R_{V}$ obtained by
minimizing residuals in the Hubble diagram using different sub-samples
and different filter combinations. As we wish to simultaneously fit
several filters at once using templates and with a single consistent
$R_{V}$, we will derive our own calibration parameters in \S
\ref{sub:Tripp-Model}.

To further complicate matters, because the SED of a SN~Ia evolves
with time, not only is $R_{x}^{yz}$ a function of $R_{V}$, it is
also a function of time and to a smaller extent, $E\left(B-V\right)$
itself. As a result, the 4th term on the right-hand-side of equation
(\ref{eq:DM_EBV}) should, strictly speaking, be a function of $E(B-V)$,
and $t$. However, computing these behaviors is computationally expensive
and the magnitude of these effects are much smaller than the intrinsic
dispersions in the light-curve templates (see \S \ref{sub:Template-Errors}).
We therefore treat $R_{x}^{BV}$ solely as a function of $R_{V}$
and compute its value at the time of maximum for a typical $E(B-V)=0.1$.

This method was first used by \citet{1999AJ....118.1766P} and we will refer
to it as the ``Color Excess'' model.

\subsection{Reddening-Free Magnitudes\label{sub:Reddening-Free-Magnitudes}}

There are at least two problems with treating $f\left(y-z\right)$
as reddening due to dust: 1) determining which SNe are unreddened
requires prior knowledge of the source of the reddening and 2) obtaining
a \emph{truly} unreddened sample is extremely unlikely. As we will
show, when one is interested only in distance,
the extra step of isolating a subsample of SNe~Ia believed to
be unreddened is an unnecessary one that only serves to introduce
a possible source of systematic error.  However, if one is interested
in the properties of the SN~Ia, such as intrinsic colors, extinction,
or the possibility of a varying reddening law, then the color excess
method is required.  The reddening model also has the advantage of
easily combining all filters into one model.

Despite this difficulty in separating what is reddening and what is
intrinsic color variation, for the cosmologist these are simply nuisance
variables. This has led several authors to simply combine the two
effects into one generalized color term and marginalize over it when
solving for the cosmological parameters 
\citep{2006A&A...447...31A,2009ApJ...700.1097H}.
This is a sensible way to proceed and is a consequence
of using reddening-free magnitudes 
\citep{1963bad..book..204J,2009ApJ...704.1036F}.
A reddening free magnitude is defined as
\begin{equation}
   w_{x}^{yz}\equiv m_{x}-R_{x}^{yz}\left(m_{y}-m_{z}\right)
   \label{eq:w_defined}
\end{equation}
where $m_{x}$, $m_{y}$ and $m_{z}$ are magnitudes observed through
filters $x$, $y$ and $z$. It is easy to show that any reddening
that obeys equation (\ref{eq:red_law}) will leave $w_{x}^{yz}$ invariant.
Therefore, by calibrating SNe~Ia using reddening-free magnitudes, we
obtain a standardizable candle that requires no knowledge of their
intrinsic color properties and no need to generate a subsample of
``unreddened'' SNe. Using the same parameterization as before,
we can define the absolute reddening-free magnitude of a SN~Ia as
\begin{equation}
W_{x}^{yz}=a_{x}^{yz}+b_{x}^{yz}\left(\dm-1.1\right)\label{eq:W_calib}
\end{equation}
where now $a_{x}^{yz}$ and $b_{x}^{yz}$ are determined from the
low-$z$ sample of SNe~Ia without the need to first determine intrinsic
colors and perform a reddening correction. The model for the observed
light curve for a SN~Ia is now\begin{equation}
m_{x}\left(t\right)=T_{x}\left(t-t_{max};\dm\right)+a_{x}^{yz}+b_{x}^{yz}\left(\dm-1.1\right)+R_{x}^{yz}\left(m_{y}-m_{z}\right)+\mu\label{eq:model_w}\end{equation}
Again, $R_{x}^{yz}$ can be left as a free parameter, determined by
the low-$z$ calibrating sample. Furthermore, if the reddening of SNe~Ia 
is truly due to dust alone, then solving for $R_{x}^{yz}$ can,
in principle, constrain the properties of the dust grains.
This type of standardization was first used by \citep{1998A&A...331..815T}
and so we refer to it as the Tripp method. 

Comparing equations (\ref{eq:DM_EBV}) and (\ref{eq:model_w}) reveals
that they are mathematically equivalent, with the realization that
$a_{x}^{BV}=a_{x}-R_{x}^{BV}a_{BV}$ and $b_{x}^{BV}=b_{x}-R_{x}^{BV}b_{BV}$.
The only difference is that the method of \S \ref{sub:red_param}
has the extra step of identifying the unreddened sample and using
it to fit for $a_{BV}$ and $b_{BV}$. These two parameters will have
formal errors, which must then be carried through as systematic errors
in the determination of the distance modulus. For example, if $a_{BV}$
were in error by $+0.01$ magnitudes, then all color excesses would
be in error by $+0.01$ and therefore the distance modulus would be
in error by $-0.01R_{x}^{BV}$.

The use of $R_{x}^{yz}$ as the coefficient to the color term should
not be taken as an endorsement of the idea that reddening is the sole
cause of the color-luminosity correction in SNe~Ia. Unlike \S
\ref{sub:red_param}, we make no assumptions on the relationships
of the different $R_{x}^{yz}$, nor impose any priors on their values.
Indeed, negative values are possible which would be considered unphysical
if interpreted as reddening coefficients. We simply hold to this notation
to emphasize its motivation (see equation (\ref{eq:w_defined})).

\subsection{SN~Ia Calibration}

We now proceed to use our sample of low-$z$ SN~Ia in order to
calibrate the parameters from \S \ref{sub:red_param} and 
\S \ref{sub:Reddening-Free-Magnitudes}.
This has been done previously in F10. In this case,
however, we take a somewhat different approach. First, we use all
the SNe in the sample with redshift greater than 0.01, including those
whose peak brightnesses were \emph{not} observed. Second, we use the
maximum brightnesses and $\dm$ values as derived from template
fits alone (i.e., we do not mix template and spline fits as was done
in F10). Third, in the case of the Color Excess model,
we fit all the filters simultaneously, deriving a single best-fit
value for the reddening coefficient $R_{V}$. Finally, we treat the
extinction of each SN~Ia in the Color Excess model as a nuisance parameter
to be determined as part of the fitting procedure. Because the Tripp
method does not assume any functional relationship between the color
coefficients of the different filters, we solve each filter combination
separately as was done in F10.

To calibrate the absolute magnitudes, we must assume a distance
modulus for each SN~Ia.  We use the values determined by F10, most of which
are determined from the CMB redshift and standard values of the cosmological
parameters:  $H_0 = 72\ {\rm km\ s^{-1}\ Mpc^{-1}}$, $\Omega_M = 0.28$, and
$\Omega_\Lambda = 0.72$ \citep{2007ApJS..170..377S}.  

In order to derive the best-fit parameters, we have chosen to use
a Markov-Chain Monte-Carlo (MCMC) approach, as it offers several advantages
over the more traditional least-squares method. Most importantly,
it is a less biased estimator of the regression parameters when significant
error is present in the control variables \citep[see][]{2007ApJ...665.1489K}.
Furthermore, MCMC allows one to use more sophisticated models of the
statistical processes that produced the data. Lastly, because the
output of the MCMC method is a set of realizations of the parameters
drawn from the posterior probability distribution (PPD), one can easily
derive confidence intervals and covariances between the parameters.
These covariances are particularly useful when estimating the errors
in distances derived using either method. Details of the MCMC method
and the specific models we use to fit the data are presented in Appendix
\ref{sec:MCMC}.

\subsubsection{Tripp Model\label{sub:Tripp-Model}}

With the Tripp model, we are simply doing regression in 3 dimensions,
so we describe this model first. We choose 3 filters $xyz$ from which
we can derive reddening-free magnitudes: $m_{x}-R_{x}^{yz}\left(m_{y}-m_{z}\right)$.
We then fit the model \begin{equation}
m_{x}=M_{x}^{yz}+b_{x}^{yz}\left(\dm-1.1\right)+R_{x}^{yz}\left(m_{y}-m_{z}\right)+\mu\left(z\right),
\label{eq:tripp_model}\end{equation}
solving for the $M_{x}^{yz}$, $b_{x}^{yz}$, and $R_{x}^{yz}$.   Each SN~Ia
is fit with light-curve templates, from which we can directly determine the
maximum light in each filter: $m_x$, $m_y$, and $m_z$.  The colors
in equation (\ref{eq:tripp_model}) are therefore pseudo-colors $y_{max}-z_{max}$
instead of observed colors at a particular epoch.  We also solve
for the intrinsic scatter in the relation, $\sigma_{SN}$ as part of the
MCMC simulations (see Appendix \ref{sec:MCMC}).  

\begin{figure}
\includegraphics[width=6in]{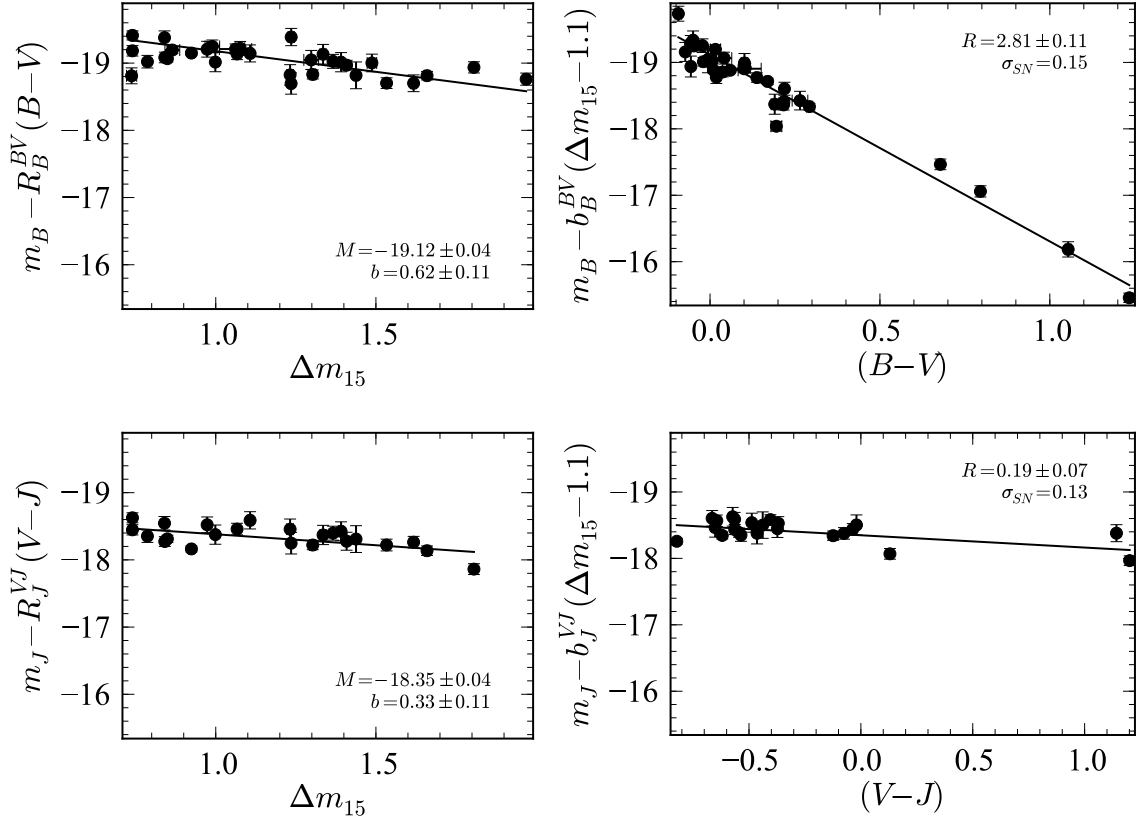}

\caption{Tripp relation derived from MCMC simulation for two filter combinations.
The first row shows filter $B$ corrected by pseudo-color $B_{max}-V_{max}$.
The second row shows filter $J$ corrected by pseudo-color $V_{max}-J_{max}$.
The left panels show the absolute magnitudes, corrected for color,
while the right-hand panels show the absolute magnitudes corrected
for $\dm$. All SNe with redshifts greater than 0.01 are
included in the fit. The fit results are labeled. \label{fig:Tripp-relation}}

\end{figure}

The results of the modeling are summarized in \tab\ \ref{tab:trip-param}.
Two sample solutions are shown in \fig\ \ref{fig:Tripp-relation}.
We find that these values are consistent with F10 though, as mentioned
earlier in \S \ref{sub:dm15_defined}, the slopes $b_{x}^{yz}$ tend to
be smaller.  This demonstrates
that the use of template fits rather than spline fits does not introduce
large systematic differences in the calibration when using
the Tripp method.  As in F10, we calibrate with two subsamples:  1) all
SNe~Ia in the Hubble flow, and 2) all SNe~Ia in the Hubble flow, but excluding
events with $B-V > 0.4$ (which excludes the two very reddened SNe~Ia
SN~2006X and SN~2005A, as well as the fast-declining SNe~Ia SN~2005bl,
SN~2005ke, and SN~2006mr).

\subsubsection{Color Excess Model}

\label{sub:EBV_model}In this case, we proceed somewhat differently.
We wish to correct using a color excess, for example $E(B-V).$ But
any combination of two filters could be used to construct
a color excess (for instance, F10 used several combinations).
In fact, given a value of $E(B-V)$ and $R_{V}$, one can use the
CCM+O extinction law to {\em predict} any other color excess. So instead
of using observed colors, we parameterize the color-dependence as
a single color excess for each SN, $E(B-V)$, and use CCM+O to fit all
wavelengths simultaneously. Under the assumption that the color-luminosity
correlation is due to reddening by dust, this allows us to solve for
one ``reddening law'', $R_{V}$ for the entire sample instead
of treating each color coefficient separately, as we did in \S
\ref{sub:Tripp-Model}. For a given value of $R_{V}$, we can derive
the ratio of total-to-selective absorption for any CSP filter $x$,
$R_{x}$, by using its filter function, the SED of a typical SN~Ia
\citep{2007ApJ...663.1187H}, and an extinction law 
$A\left(\lambda;R_{V},E(B-V)\right)$
derived from CCM+O. We do this by using the following formula:
\begin{equation}
R_{x}\left(R_{V},E(B-V)\right)=
\frac{-2.5}{E(B-V)}\log_{10}
\left(\frac{\int\Phi\left(\lambda\right)R_{x}\left(\lambda\right)A\left(\lambda;R_{V},E(B-V)\right)d\lambda}
{\int\Phi\left(\lambda\right)R_{x}\left(\lambda\right)d\lambda}\right)
\label{eq:R_X}.
\end{equation}  We also solve for the intrinsic dispersion, $\sigma_{SN}$,
for all filters combined.

In F10, the color excess was computed by isolating an un-reddened
sample of SNe in order to determine both the intrinsic colors as a
function of $\dm$ as well as establishing a Lira Law 
\citep{1996MsT..........3L}
for the CSP filters. In our MCMC model, we choose to instead
treat each $E(B-V)$ as a parameter to be determined. We cannot assume
a uniform prior on these values due to the degeneracy between the
extinction and the absolute magnitudes $M_{x}\left(0\right)$. We
therefore must employ a prior on the extinction in order to penalize
arbitrarily large values of $E(B-V)$. We do this in two different
ways.

First, we can use the ``unreddened'' subsample from F10 and assign
these SNe zero extinction, with some intrinsic scatter $\sigma_{c}$
and the rest are allowed to have non-zero extinction. We shall refer
to this as the ``blue prior''.  Second, we can
forget the ``unreddened'' sample and assume a prior for $E(B-V)$
peaked at zero, with an intrinsic scatter of $\sigma_{c}$ and with a tail
to higher extinctions with characteristic length $\tau$ . This is
motivated by the work of \citet{1998ApJ...502..177H}, who modeled
the likely extinction disribution in host disk galaxies, and the
later analysis of \citet{2007ApJ...659..122J} who found
that the colors of SNe~Ia follow such a distribution 
(see Figure 6 of \citet{2007ApJ...659..122J} and Figure 17 of
\citet{2009ApJS..185...32K}).  We shall refer to this as the
``Jha prior''.
Assigning this prior to each SN~Ia, the MCMC method should converge
such that $E(B-V)$ = 0 for the bluest SNe.

Using these two MCMC models, we fit the calibration parameters once
with all normal SNe~Ia ($\dm < 1.7$) in the sample then again
with the two very red SNe (SN~2005A
and SN~2006X) removed. This gives us 4 sets of calibration parameters
that allow us determine the systematic effects of the red SNe and of
the assumed extinction prior. 

\begin{figure}
\includegraphics[width=6in]{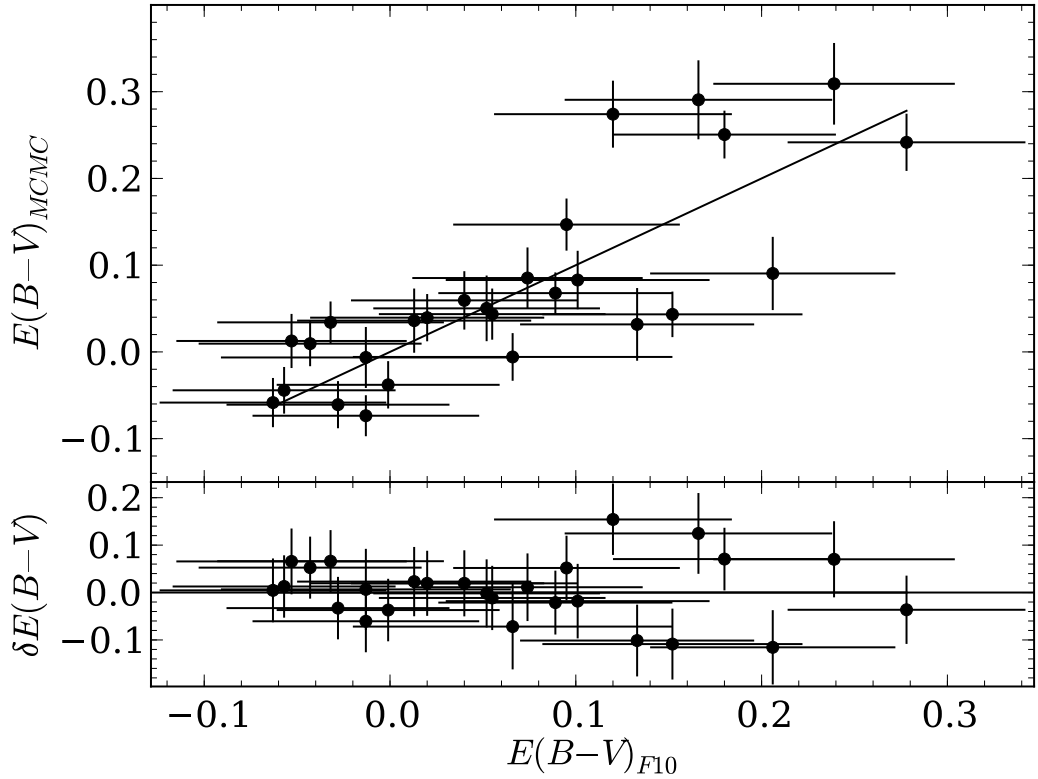}

\caption{Comparison of values of $E(B-V)$ derived by F10 and determined by
the MCMC blue prior model, excluding the two red SNe. The solid line indicates
$E(B-V)_{F10}=E(B-V)_{MCMC}$ and is not a fit to the data. The bottom
panel shows the residuals about the solid line.\label{fig:EBVcompF10}}

\end{figure}

\fig\ \ref{fig:EBVcompF10} compares the values of $E(B-V)$ derived through 
the MCMC model using the blue prior to those of F10.  There is a good deal of scatter
somewhat above what would be expected from the errors alone.  However,
there is a clear trend that the most reddened objects in F10 are
the most reddened objects in the MCMC analysis.  The excess scatter
is due to a systematic trend with \dm\ which we describe below.

\begin{figure}
\includegraphics[width=3in]{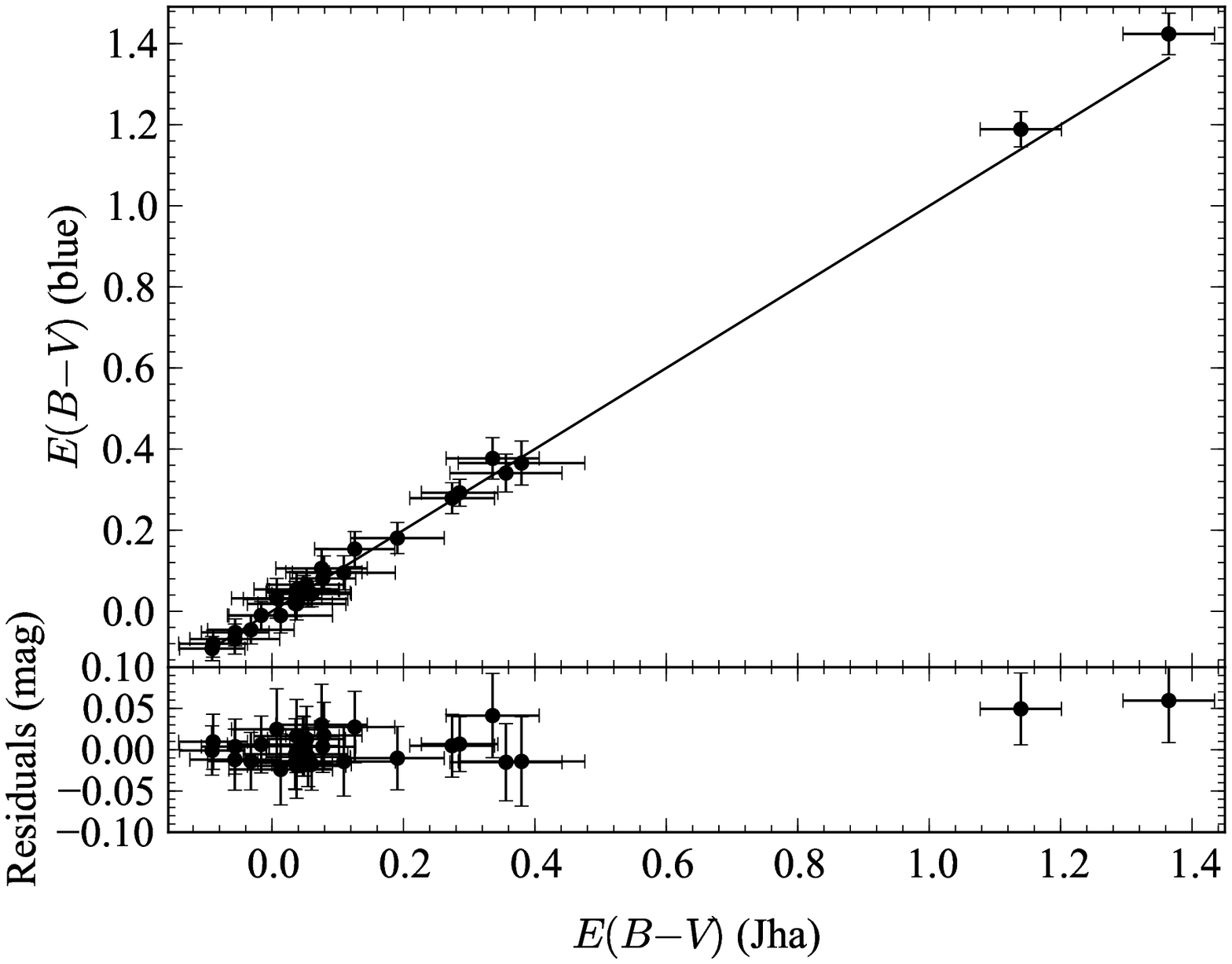}\includegraphics[width=3in]{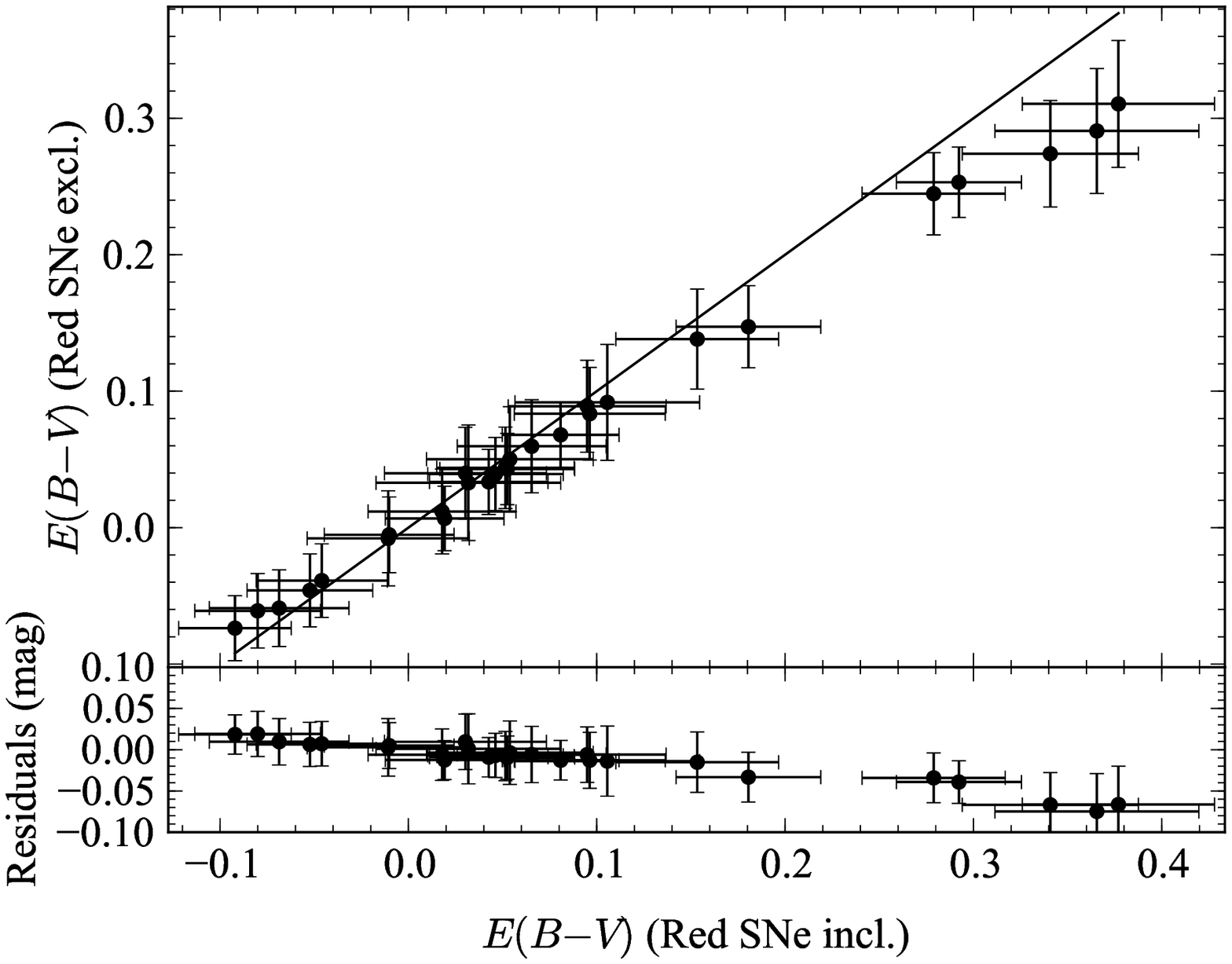}

\caption{Comparison of values of $E(B-V)$ derived by MCMC.  Left: two different
priors (the blue prior and Jha prior) are used.  Right:  two different sub-samples of SN~Ia (including and
excluding the two red SNe~Ia) are used with the blue prior. In both
figures, the solid line shows one-to-one correspondence.\label{fig:EBVcomparisons}}

\end{figure}

The left panel of \fig\ \ref{fig:EBVcomparisons} shows the difference between the
values of $E(B-V)$ derived using the two different extinction priors.
As one can clearly see, there is no large systematic difference between
the two sets. However, whether we include the two red SNe \emph{does}
have a systematic effect, as can be seen in the
right panel of \fig\ \ref{fig:EBVcomparisons},
where we use the same extinction prior.
This is in agreement with the results of F10, who found that the two
red SNe~Ia seem to follow a different reddening law than the bluer
SNe~Ia. Due to their very red colors, SN~2005A and SN~2006X
have a large pull on the values of $R^{BV}_X$ in equation (\ref{eq:DM_EBV}),
favoring smaller values.  The MCMC simulation then responds by
modifying the values of $E(B-V)$ in such a way that the
average color correction is preserved:  the redder objects have more
color excess to compensate the smaller $R^{BV}_X$.
This can also be seen in \tab\ \ref{tab:EBV-global}
where we list the wavelength-independent calibration parameters. The red SNe~Ia drive
$R_{V}$ to lower values, however the bluer SNe~Ia prefer a larger
$R_{V}$. As a consequence, including the red SNe also increases the 
derived intrinsic
dispersion in the SNe~Ia.

\begin{figure}
\includegraphics[width=6in]{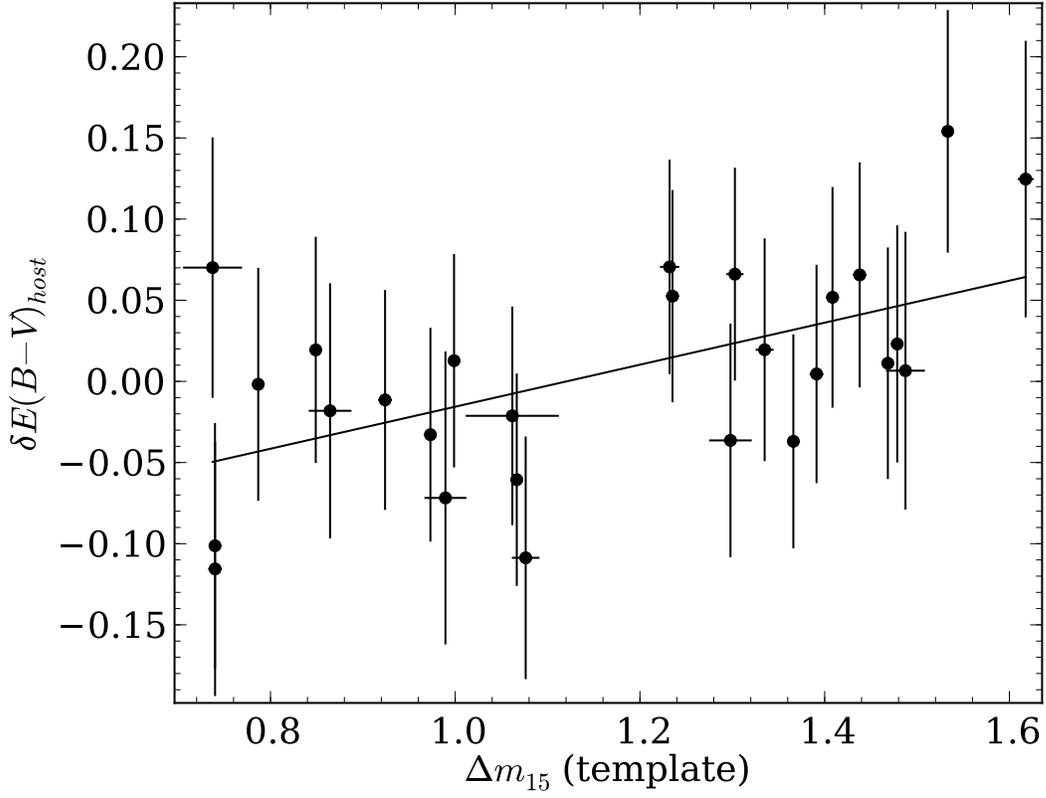}

\caption{Difference between the $E(B-V)$ values derived by F10 and those determined
through MCMC modeling excluding the red sample as a function of $\dm$.
The solid line is a fit to the trend: $E(B-V)_{MCMC}-E(B-V)_{F10}=-0.15+0.13\dm$.
\label{fig:dEBV_dm15}}

\end{figure}

The extinction prior and inclusion of the red SNe~Ia has a smaller
effect on the filter-specific calibration parameters, which all agree
to within the statistical errors. A marked difference between these
results and those of F10 is the difference in the slopes, $b_{x}$.
In F10, the slopes were systematically higher and became smaller with
redder filter. 
The reason for the smaller slopes is due to two systematic effects.
First, as shown earlier in \fig\ \ref{fig:compare_dm15s}, there is a
systematic difference between $\dm(B)$ and $\dm$ such that
$\dm(B) \sim 0.9\dm$, and so the $\dm$ correction factor, $b_x$, 
will tend to be smaller.  Second, there is also a systematic
difference between our estimates of $E(B-V)$ and those from
F10 as a function of
$\dm$. These are shown in \fig\ \ref{fig:dEBV_dm15} and
clearly indicate a systematic difference correlated with $\dm$
in the sense that larger $\dm$ produces larger $E(B-V)$
than estimated in F10. In this way, the $\dm$ dependence
is partly absorbed into the $E(B-V)$ term of equation (\ref{eq:DM_EBV}),
resulting again in smaller $b_{x}$.  
The reason for this systematic difference in $E(B-V)$ is that we
have excluded SN~2006gt from our sample of ``unreddened'' SNe~Ia.  This
one object had a relatively large $\dm \sim 1.7$ and very red colors ($B-V \sim 0.2$)
and therefore tended to make the SNe~Ia intrinsically redder for larger $\dm$.

\begin{figure}
\includegraphics[width=6in]{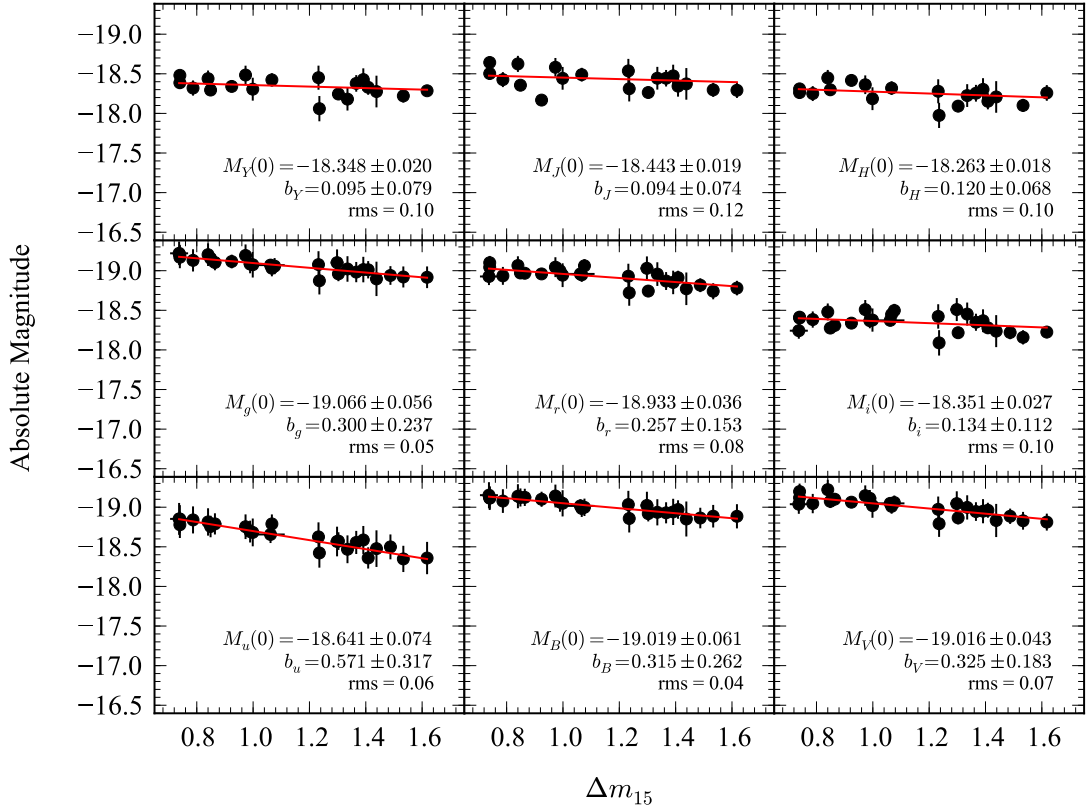}

\caption{Luminosity-\dm\  relation for the Color Excess model
derived from MCMC fitting, excluding the two red
SNe and three fast-declining SNe. Each panel corresponds one of the
CSP filters. The distance moduli from F10 are used to compute the
absolute magnitudes of each SN. SNooPy fits are used to compute $\dm$.
The SNe are reddening-corrected using the values of $E(B-V)$ and
$R_{V}$ from the MCMC run. The solid lines show the fit to $M_{x}$
and $b_{x}$, which are labeled with uncertainties in each panel.
\label{fig:reddening-relation}}

\end{figure}

The calibration parameters for the Color Excess model are given in 
\tab\ \ref{tab:EBV-params}.  For each parameter we show values derived 
with and without the red
SNe, and also with the two different $E(B-V)$ priors. Due to the
fact that there does seem to be something different about including
these extremely red events, we recommend using the calibration which
was derived \emph{without} them. This subsample most closely resembles
calibration 6 in \tab\ 9 of F10.  The fits to the absolute magnitude-
\dm\ relation for this subsample are shown in \fig\ 
\ref{fig:reddening-relation}.

\subsection{Systematic Error Budget}

While fitting light curves with any  least-squares method%
\footnote{We use the Lavenberg-Marquardt least squares algorithm, as implemented in the 
python package scipy.} supplies 
us with reliable statistical
errors on the light-curve parameters, care must be taken when computing
the error in the absolute distance to a single galaxy hosting an individual
(or group of) SN~Ia \citep{2008AJ....136.1482S,Stritzinger2010}.
There are 4 kinds of systematics that need to be taken into account
in order to report accurate error estimates on SN~Ia-derived distances:
\begin{enumerate}
\item The Hubble law was used to determine the absolute magnitudes of the
SN~Ia sample, and so the assumed Hubble constant 
($H_{0}=72~{\rm km}\cdot {\rm s}^{-1}\cdot {\rm Mpc}^{-1}$)
sets the scale of all derived distances. Currently, there 
has been a marked improvement in the error budget for $H_0$,
having been reduced from $\pm 10$\% \citep{2001ApJ...553...47F}
to $\pm 5$\% \citep{2009ApJ...699..539R,FreedmanMadore2010}.
Any further improvement of this error
in $H_{0}$ will directly benefit SN~Ia-derived distances.
\item The uncertainties the calibration parameters $M_{X}$, $b_{X}$,
and $R_{V}$ introduce systematic errors in the distance. The magnitude
of these errors will depend on which filters are used and their relative
weights. There are also significant covariances between the calibration
parameters, particularly between the $M_{X}$ and $R_{V}$. For this
reason, computing the systematic error is best done using Monte-Carlo
techniques, drawing from the posterior probability distribution output
by the MCMC run. We have included routines to do this in SNooPy.
\item After the $\dm$ and color-dependent corrections have been
performed, there remains an intrinsic dispersion in the SN~Ia distances
that is not explained by measurement error. This extra dispersion
$\sigma_{SN}$ should be added in quadrature to the other systematic
errors for a single event. However, if multiple events are used to
determine the distance to a galaxy (see e.g., \citet{Stritzinger2010}), 
then these errors
should reduce by $\sqrt{N}$%
\footnote{There is some evidence
that the remaining residuals correlate with host galaxy
properties \citep{2008ApJ...685..752G,2010MNRAS.tmp..755S,2010arXiv1005.4687L}. 
In this case,
one could argue the $\sigma_{SN}$ errors for a single galaxy would not reduce as
$\sqrt{N}$. However, it is just as likely that the residuals are
a function of the progenitor's environment, which could vary greatly
throughout a galaxy.%
}.
\item If the SN~Ia was observed in a photometric system different than the
CSP, or if the redshift of the object is sufficiently large to require
cross-band K-corrections, then the errors in the zero-points of each
filter must be included. For the CSP filters, this amounts to approximately
0.02 mag in the distance modulus (1\% in distance). These must be
added in quadrature with the errors of any other photometric system
used to observe the SN~Ia.
\end{enumerate}
Which of these systematics should be included depends greatly on the
application. For instance, when using SNe~Ia for cosmology, the error
in the Hubble constant drops out, since we are only interested in
the \emph{relative} distances between the SN~Ia. However, if the absolute
distance to a galaxy is the quantity of interest, then all these errors
must be taken into account.

\subsection{Hubble Diagram and Host Distances}

\begin{figure}
\includegraphics[width=6in]{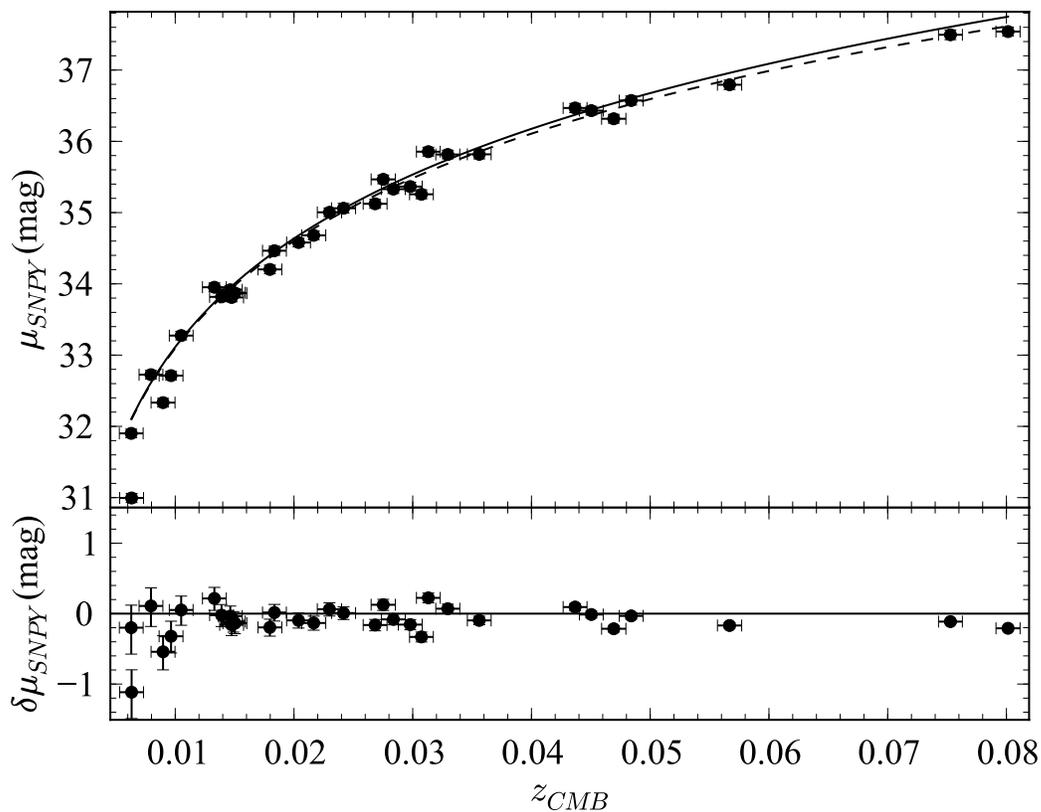}

\caption{Hubble Diagram constructed from the observed CMB velocities and SNooPy-derived
distance moduli, using the full $uBVgriYJH$ filter set. The points
are individual SN~Ia with error-bars from the fit. The solid line
shows the standard cosmology redshift-distance relation, while the
dashed line shows the simple Hubble Law for $H_{o}=72\ {\rm km}\cdot {\rm s}^{-1}\cdot {\rm Mpc}^{-1}$.
\label{fig:Hubble_all}}

\end{figure}

Now that we have a calibration from the low-$z$ sample, we can fit the
full $uBVgriYJH$ light curves of all 34 SNe in our sample (we do not
use the two SN~Ia that have $\dm>1.7$). To illustrate this,
we construct a Hubble diagram using the measured CMB velocities from
NED, and the distance moduli yielded by the $E(B-V)$ model with
the calibration parameters derived in \S \ref{sub:EBV_model}.  \tab\ 
\ref{tab:LCparams} lists the parameters derived from the SNooPy fits, the
last column showing the distance modulii.
We choose the calibration for which we excluded the two red SNe~Ia
and used the blue subsample to anchor the colors, as this contains
the fewest nuisance parameters. The resulting Hubble diagram is shown
in the top panel of \fig\ \ref{fig:Hubble_all}. The points are the
individual SNe~Ia. The solid line shows the standard cosmology 
($H_{o}=72 {\rm km}\cdot {\rm s}^{-1}\cdot {\rm Mpc}^{-1}$,
$\Omega_{m}=0.28$, $\Omega_{\Lambda}=0.72$) used in F10 while the
dashed line shows the simple Hubble law $v=H_{o}D$ for comparison.
The solid line is not a fit to the data, as this would be completely
circular, since the cosmology was assumed to derive the calibration
parameters in the first place. The residuals between the SNooPy-derived
distances and standard cosmology are shown in the bottom panel of
\fig\ \ref{fig:Hubble_all}. As expected, the dispersion increases
at lower redshifts, where peculiar velocities exceed the Hubble flow
velocity.  SN~2006X, in particular, stands out at the low-$z$ end
of the Hubble diagram.  Its residual is larger due to peculiar velocities,
but also because we chose the calibration that excluded it and SN~2005A.

While the lowest-redshift SNe~Ia ($z<0.01$) cannot be used to calibrate
the Tripp or Color Excess relations, nor can they be used to constrain
cosmological models, they are still of interest, as they serve as
distance indicators to their host galaxies. As such, SNe~Ia can serve
as useful standard candles to those galaxies that lie in the ``gap''
between the furthest measured Cepheids  and the distance at which the Hubble
flow can be considered to be in excess of any expected peculiar velocities.
The SNe~Ia (and hosts) that have $z<0.01$ are: SN~2005W (NGC 691),
SN~2005am (NGC 2811), SN~2005ke (NGC 1371), SN~2006D (MRK 1337),
SN~2006mr (Fornax A), and SN~2007af (NGC 5584). 

Of these low-$z$ objects, we can determine distances to those for which 
$\dm < 1.7$.  According to NED, NGC~691 has a Tully-Fisher distance of $\mu = 32.71\pm0.40$ mag
\citep{2008ApJ...686.1523T}.  This compares very well with the SNooPy distances of 
$\mu = 32.73\pm 0.13$ mag using SN~2005W.  We could not find a velocity-independent distance for 
NGC~2811 with which to compare our distance of $\mu = 32.33 \pm 0.18$ mag, however it is 
reasonably close to the Virgo-corrected $\mu = 32.52 \pm 0.15$ mag reported by NED.
The distance to MRK~1337 has been measured by other authors 
\citep{2009ApJ...704..629M,2008ApJ...689..377W} who derive a distance of
$\mu = 32.72 \pm 0.06$ mag using NIR observations of SN~2006D.  We therefore 
contribute an independent distance of $\mu = 32.71 \pm 0.14$ to SN~2006D, 
which agrees very well.  The treatment of the distance of Fornax A using
SNooPy is 
discussed in a separate, dedicated paper \citep{Stritzinger2010}.   Finally,
the distance to NGC~5584 has a Tully-Fisher distance of $\mu = 31.48 \pm 0.52$ mag
\citep{2008ApJ...686.1523T}.  While quite different from the SNooPy distance of
$\mu = 31.90 \pm 0.11$, it is well within the error and agrees well with the
Virgo-corrected $\mu = 32.10\pm0.15$ mag distance.

\section{Conclusions}
\label{sec:conclusions}
In this paper, we have presented SNooPy, a SN~Ia light-curve template
generator and method appropriate for determining distances to nearby
galaxies. These templates have also been used to interpolate the peak
rest-frame $i$-band fluxes of a sample of high-$z$ SNe,
allowing the first $i$-band Hubble diagram out to redshifts of 0.7. 
To our knowledge, SNooPy is the only fitter that can
simultaneously fit SN~Ia light curves from $u$ band all the way to
the NIR $H$ band.  It therefore has the advantage of a very large
wavelength coverage.

This purpose of this paper was mainly to present the methodology used
by the CSP to fit light-curves and provide a calibration that is more
consistent with the way SNooPy determines distances.  The details of
the light-curve templates and the calibration parameters will no doubt
change as more SNe~Ia are added to the low-z sample.  In particular, we
caution the reader that the small number (2) of significantly reddened
objects leads to significantly different results when they are included
in the calibration.  We therefore recommend that these calibrations {\em not}
be used and present them only to illustrate the effects of including
these highly reddened SNe~Ia.  Indeed, objects such as these would be
selected against in high-redshift surveys.  The mere fact that the CSP
SNe were selected in a very different way (targeted search) than the
blind high-z surveys will lead to some systematic biases.  These systematics
will be further investigated when larger numbers of CSP SNe~Ia are available.

The construction of the light-curve templates, which represent some average
behavior of a sample of SNe~Ia, has also revealed that at longer wavelengths,
there is a marked increase in the SN-to-SN variation in light-curve
behavior, even at a given decline rate. Tentatively, it also seems
that the peak of the light curves show the least dispersion as standardizable
candles. This poses a challenge for the observer, as SN events are
typically ``triggered'' in the optical bands and the light curves
peak earlier in the NIR. It is therefore often necessary to use template
light curves to extrapolate the peak magnitude. On one hand, the
increased dispersion in light-curve behavior will make this extrapolation
more uncertain in the NIR. At the same time, these variations hint
that another light-curve parameter might be at work and that this
parameter may be correlated with residuals in the 
Hubble diagram. If this turns
out to be the case, the inclusion of NIR light curves will deliver
a significant advantage to SN~Ia cosmology.

We have also presented a least-squares method to determine distances
to SN~Ia by simultaneously fitting any combination of $uBVgriYJH$
photometry. This method gives distance moduli with an rms scatter
of about 0.06 magnitudes (3\% in distance). However, added to this
small dispersion are the various systematic errors. These include
the use of the Hubble constant to determine the distances to the calibrating
sample, the formal uncertainties in the calibration parameters, the
intrinsic dispersion in the luminosities of SNe~Ia, and errors in the
photometric zero-points.

Reduction of these systematic errors will improve the use of SNe~Ia
as standard candles. There is work being done to reduce the error
in the Hubble constant to $\pm2\%$ \citep{FreedmanMadore2010}, which
will, by extension, greatly improve SN~Ia distances. In this paper,
we used approximately one third of the CSP low-$z$ sample. Including
the entire sample will reduce the errors in the calibrating constants,
thereby reducing the systematic errors. The intrinsic dispersion in
the SN~Ia luminosities, $\sigma_{SN}$, will not improve with a larger
calibrating sample. 
Lastly, the CSP is currently
working on directly scanning the filter, CCD and telescope throughputs
in order to reduce the uncertainty in the photometric zero-points
as well as improving S- and K-corrections. 

Using SNooPy, we have fit the distances of several SNe~Ia that are
not in the Hubble flow. Comparison of these distances with other
methods gives good agreement in general, particularly when comparing
the same SN~Ia distance to SN~2006D (MRK~1337).

\acknowledgements{}
We acknowledge the National Science Foundation (NSF) through
grant AST03-06969 for support of the low-$z$ component of the CSP.
The Dark Cosmology Centre is funded by the Danish NSF. 
We have made use of the NASA/IPAC Extragalactic Database (NED) which is
operated by the Jet Propulsion Laboratory, California Institute of Technology,
under contract with the National Aeronautics and Space Administration.

\appendix{}

\section{CSP Photometry\label{sec:CSP-Photometry}}

The CSP photometry has been published in natural magnitudes in order
to simplify their incorporation in other photometric systems. Here,
we briefly explain the natural system for the benefit of those who
would wish to combine our data with theirs. 

Because the filters used on a particular instrument and telescope
are not perfect matches to the filters used to establish a photometric
system of standards, such as those presented in \citet{1992AJ....104..340L}
and \citet{2002AJ....123.2121S}, color terms are needed to convert
observed instrumental magnitudes to standard magnitudes. This is typically
done by observing a sequence of standard stars with a spread in colors
and comparing the instrumental magnitudes to the published standard
magnitudes, fitting a formula of the type $A^{\prime}=A+ct_{A}\left(B-C\right)$
where $A^{\prime}$ is the magnitude in filter $A$ of a standard star
in the standard system, $A$, $B$, and $C$ are instrumental magnitudes
through filters $A$, $B$, and $C$, and $ct_{A}$ is the color term for filter
$A$. The CSP has determined their color terms and published them in
Hamuy et al (2006), which we reproduce here for convenience:\begin{eqnarray*}
B^{\prime} & = & B+0.060(0.013)\left(B-V\right)\\
V^{\prime} & = & V-0.057(0.013)\left(V-i\right)\\
u^{\prime} & = & u+0.051(0.017)\left(u-g\right)\\
g^{\prime} & = & g-0.017(0.009)\left(g-r\right)\\
r^{\prime} & = & r-0.019(0.017)\left(r-i\right)\\
i^{\prime} & = & i-0.007(0.017)\left(r-i\right)\end{eqnarray*}
The problem is that these color terms are determined using stellar
SEDs. SN~Ia, in contrast, have broad absorption features that vary
with time, and so applying these equations to the instrumental SN
Ia to obtain standard magnitudes is not valid. This is the reason
that different telescopes observing the same SN~Ia end up with significantly
different magnitudes, even though both data sets have been converted
to a ``standard system''. The solution to this problem is to publish
natural photometry along with the filter functions used to make the
observations. This is done in the following manner.

First, the equations above are used in reverse to convert the standard
magnitudes of \citet{2002AJ....123.2121S}  and \citet{1992AJ....104..340L}
into the CSP natural
system. Then these new magnitudes along with observations of the standard
stars are used to determine the zero-points for the observations of
the SN~Ia. In this way, the natural system magnitude through filter
A are equivalent to
\begin{equation}
m_{A}=-2.5\log_{10}\left(\frac{\int F_{A}\left(\lambda\right)\Phi_{SN}\left(\lambda\right)\lambda d\lambda}
{\int F_{A}\left(\lambda\right)\Phi_{std}\left(\lambda\right)\lambda d\lambda}\right)
\end{equation}
where $F_{A}$ is the filter function (including filter transmission,
CCD and telescope response functions, and atmospheric extinction)
that corresponds to the telescope that actually made the observations,
$\Phi_{SN}$ is the SED of the SN~Ia that was observed, and $\Phi_{std}$
is an average SED of the spectral standards used to determine the
zero-point. If another telescope observes the same event through a
different filter B and the magnitudes are published in its standard
system, then an S-correction is simple to determine:
\begin{equation}
S_{BA}\left(t\right)=-2.5\log_{10}
   \left(\frac{\int F_{A}\left(\lambda\right)\Phi_{SN}
   \left(\lambda\right)\lambda d\lambda\times\int F_{B}
   \left(\lambda\right)\Phi_{std}\left(\lambda\right)\lambda d\lambda}
   {\int F_{B}\left(\lambda\right)\Phi_{SN}\left(\lambda\right)\lambda 
   d\lambda\times\int F_{A}\left(\lambda\right)\Phi_{std}\left(\lambda\right)\lambda d\lambda}\right)
\end{equation}
where $S_{BA}\left(t\right)$ is an S-correction that converts a magnitude
from system B into system A: $m_{A}\left(t\right)=m_{B}\left(t\right)+S_{BA}\left(t\right)$.

SNooPy can perform this transformation automatically by using the
\citet{2007ApJ...663.1187H} SED templates if 1) the input photometry
is in a natural system like that described above, and 2) the filter
functions of the observed photometric system are supplied. In effect,
when both systems are natural, then the S-correction can be considered
to be a cross-band K-correction at low redshift.

\section{MCMC Modeling}

\label{sec:MCMC}The use of Markov-Chain Monte-Carlo (MCMC) simulations
to model data is rapidly growing in popularity in the astronomical
community. It has many advantages over the more commonly used $\chi^{2}$
analysis. Most importantly, it allows one to use more realistic statistical
models for the data and priors, rather than assuming all errors are
normally distributed. Second, the result of the MCMC simulation is a set
of parameters drawn from the posterior probability distribution, from
which one can easily determine expectation values, modes, variances
and co-variances of the parameters of interest. 

Readers who are unfamiliar with MCMC may wish to read the pymc 
documentation\footnote{Available at the following URL: http://code.google.com/p/pymc/%
}. While not mathematically rigorous, it provides an excellent overview
of the more technical aspects of the method and its terminology. Pymc
is a Python module that greatly simplifies the task of setting up
and running MCMC simulations and was used to run the simulations in
this paper. The code defining the statistical model is available at
the CSP website.

MCMC is a method based on Bayesian statistics. Bayes' theorem states
that the probability distribution of the parameters $\theta$ of a model,
given the observed data $D$, is given by\begin{equation}
p\left(\theta|D\right)=\frac{p\left(D|\theta\right)p\left(\theta\right)}{p\left(D\right)}\label{eq:Bayes}\end{equation}
where $p\left(D|\theta\right)$ is the probability that we observe
$D$, given $\theta$, $p\left(\theta\right)$ is the prior distribution
of the parameters, and $p\left(D\right)=\int p\left(D|\theta\right)p\left(\theta\right)d\theta$.
The functional form of $p\left(D|\theta\right)$ is straightforward,
as is $p\left(\theta\right)$. However, the set of parameters $\theta$
will likely contain nuisance parameters over which we wish to marginalize.
Performing integrals of the right-hand side of equation (\ref{eq:Bayes})
analytically can only be done for the simplest PDFs and priors. Assuming
normal distributions, for instance, leads to the well-known $\chi^{2}$
statistic and the method of least squares. For anything more complex,
one must numerically integrate equation (\ref{eq:Bayes}) to marginalize,
compute expectation values, etc.

Markov chain Monte-Carlo is a method for dealing with the situation
where equation (\ref{eq:Bayes}) cannot be integrated analytically.
The method works by creating a Markov Chain of parameter states $\theta_{i}$.
Markov Chains have the property that the state $\theta_{i+1}$ depends
only on the previous state $\theta_{i}$. The transition from state
$\theta_{i}\rightarrow\theta_{i+1}$ is done in a probabilistic way
using equation (\ref{eq:Bayes}), hence the use of Monte-Carlo in the
method's name. The Markov chain can therefore be thought of as a quasi-random
walk through parameter space. The exact details of how the transition
is done depends on the MCMC method used. Two popular algorithms are
Gibbs sampling and the Metropolis-Hastings algorithm. Regardless of
the method, it can be shown that after a certain number of transitions,
called burn-in, the Markov chain will become stationary. From that
point on, the distribution of states in the Markov chain is equal
to $p\left(\theta|D\right)$. In other words, the states in the Markov
chain can be considered as random draws from $p\left(\theta|D\right)$.
One can therefore use the Markov chain to infer the posterior probability
distribution of all the parameters of interest. For this paper, we
use the Metropolis-Hastings algorithm.

In order to use the MCMC method, one must construct a probabilistic
process that relates the data to the model and specifies any priors
$p\left(\theta\right)$. From this, pymc can compute the likelihood.
We consider two models, one for the Tripp method and one for the Color Excess
method. The Tripp method, being simply linear regression, is the simplest
and we describe it first. We then describe the Color Excess model, which
has more complicated priors.

\subsection{Tripp Model}

Given three filters $XYZ$, the data consist of the 3 magnitudes at
maximum $m_{x}$, $m_{y}$ and $m_{z}$, the decline-rate $\dm$,
and the redshift $z$. We wish to fit the following model 
\begin{equation}
   m_{x}^{\prime}=M_{x}^{yz}+b_{x}^{yz}
   \left(\Delta^{\prime}m_{15}-1.1\right)+
   R_{x}^{yz}\left(m_{y}^{\prime}-m_{z}^{\prime}\right)
   +\mu\left(z^{\prime}\right)\label{eq:tripp_model_app}
\end{equation}
solving for the calibration parameters $M_{x}^{yz}$, $b_{x}^{yz}$,
and $R_{x}^{yz}$. The primes in this equation denote ``true''
values, in order to distinguish them from observed values. The ``true''
values are nuisance parameters and will be marginalized. We assume
that the observables are statistically related to their ``true''
values through a normal distribution:\begin{equation}
\left[m_{x},m_{y,}m_{z},\dm,z\right]\sim N_{5}\left(\left[m_{x}^{\prime},m_{y,}^{\prime}m_{z}^{\prime},\dm^{\prime},z^{\prime}\right],C\right)\label{eq:N5}\end{equation}
where $N_{5}$ is a 5-dimensional multivariate normal distribution
and $C$ is the covariance matrix. The covariances between the light-curve
parameters are provided by the least-squares fitting. The value of
$m_{x}^{\prime}$ in equation (\ref{eq:N5}) is given by equation (\ref{eq:tripp_model_app}).
We also add an extra term $\sigma_{SN}^{2}$ to the $(m_{x},m_{x})$
term of the covariance matrix $C$ that represents any intrinsic scatter
in the luminosity of SNe~Ia. We leave this as a free parameter and
so its value will be determined along with the calibration parameters.
Finally, we assign the following priors to the calibration parameters:\begin{eqnarray}
M_{x}^{yz} & \sim & U(-\infty,\infty)\label{eq:priors}\\
b_{x}^{YX} & \sim & U(-\infty,\infty)\nonumber \\
R_{x}^{yz} & \sim & U(-\infty,\infty)\nonumber \\
\sigma_{SN} & \sim & U(0,\infty)\nonumber \end{eqnarray}
where $U(x,y)$ is a uniform prior between $x$ and $y$. This completes
the statistical model. The parameters consist of the set $\theta=\left\{ M_{x}^{yz},b_{x}^{yz},R_{x}^{yz},\sigma_{SN},m_{x,i}^{\prime},m_{y,i}^{\prime},m_{z,i}^{\prime},\dm^{\prime},z^{\prime}\right\} $
and the probability of any state $\theta_{i}$ can be computed through
equations (\ref{eq:tripp_model_app}) to (\ref{eq:priors}). The Metropolis-Hastings
algorithm can then construct a Markov Chain from which we can infer
the PPD of all variables of interest.

\subsection{Color Excess Model}

This model is similar to the Tripp model. In this case, however, we
fit all the filters simultaneously. Our data therefore consists of
all 9 magnitudes at maximum, $m_{x_{j}}$, the decline-rates, $\dm$,
and the redshifts, $z$. We assume the errors in the observables are
once again normally distributed: 
\begin{equation}
   \left[m_{x_{j}},\dm,z\right]\sim N_{11}\
   \left(\left[m_{x_{j}}^{\prime},\dm^{\prime},z^{\prime}\right],C\right),
\end{equation}
where now the $m_{x_{j}}^{\prime}$ are given by
\begin{equation}
   m_{x_{j}}^{\prime}=M_{x_{j}}+b_{x_{j}}
   \left(\Delta^{\prime}m_{15}-1.1\right)+R_{x_{j}}^{BV}
   \left(R_{V}\right)E(B-V)+\mu\left(z^{\prime}\right).
\end{equation}
As discussed in \S \ref{sub:red_param}, we treat the color excesses
$E(B-V)$ as free parameters. We use two different priors on the values
of $E(B-V)$. First, we use the ``unreddened'' subsample from
F10 and assign these SNe a Gaussian prior with zero mean extinction
and some unknown scatter $\sigma_{c}$. The other SNe are then assigned
a composite prior:
\begin{equation}
   E(B-V)\sim\left\{ \begin{array}{c}
                      N(0,\sigma_{c}),\; E(B-V)<0\\
                      U(0,\infty),\; E(B-V)>0
                     \end{array}
              \right.
\end{equation}
where $N\left(0,\sigma_{c}\right)$ is a normal distribution with
zero mean and standard deviation $\sigma_{c}$. In this way, the ``unreddened''
SNe anchor the values of $E(B-V)$ and any SN with colors redder than
this sample will have $E(B-V)>0$. Any SNe with significantly bluer
colors will have $E(B-V)<0$ and cause $\sigma_{c}$ to be larger.

Second, we can avoid using an unreddened sample and instead use a
prior for all the SN~Ia that penalizes high values of $E(B-V)$. In
this case, we use the prior from \citet{2007ApJ...659..122J}, which
corresponds to the convolution of a normal distribution $N(0,\sigma_{c})$
with an exponential tail $\exp(-E(B-V)/\tau)$. We leave the variables
$\sigma_{c}$ and $\tau$ as nuisance variables to be determined during
the MCMC run. In this way, the model will search for a ``blue edge" in
the color distribution of the SNe~Ia. The bluest SNe will determine
the intrinsic colors and scatter. The redder SNe will set the scale
length of the exponential tail, $\tau$.

Our parameter set now contains 19 calibration parameters: 9 absolute
magnitudes, $M_{x_{j}}$, 9 slopes $b_{x_{j}}$, and the reddening
law $R_{V}$. Added to this are the intrinsic scatter, $\sigma_{SN}$,
the intrinsic color scatter $\sigma_{c}$, the exponential tail scale
length $\tau$ (for the Jha prior), the $N$ values of $E(B-V)_{i}$,
and all the ``true'' values of the observables. 

\bibliographystyle{apj}
\bibliography{highz_paperI}

\begin{thebibliography}{58}
\expandafter\ifx\csname natexlab\endcsname\relax\def\natexlab#1{#1}\fi
\bibitem[{{Astier} {et~al.}(2006)}]{2006A&A...447...31A}{Astier}, P., {et~al.} 2006, \aap, 447, 31

\bibitem[{{Bohlin}(2007)}]{2007ASPC..364..315B}{Bohlin}, R.~C.  2007, The Future of Photometric, Spectrophotometric and PolarimetricStandardization (ASP Conf. Ser. 364), ed. C.~{Sterken} (San Francisco, CA:  ASP), 315

\bibitem[{{Cardelli} {et~al.}(1989)}]{1989ApJ...345..245C}{Cardelli}, J.~A., {Clayton}, G.~C., \&  {Mathis}, J.~S.  1989, \apj, 345, 245

\bibitem[{{Cleveland} {et~al.}(1991)}]{Cleveland1991}{Cleveland}, W.~S., {Grosse}, E., \&  {Shyu}, W.~M.  1991, in Statistical Modelsin S, ed. J.~M. {Chambers} \& T.~J. {Hastie} (Pacific Grove: Wadsworth \&Brooks)

\bibitem[{{Conley} {et~al.}(2008)}]{2008ApJ...681..482C}{Conley}, A., {et~al.} 2008, \apj, 681, 482

\bibitem[{{Contreras} {et~al.}(2010)}]{2010AJ....139..519C}{Contreras}, C., {et~al.} 2010, \aj, 139, 519

\bibitem[{Dierckx(1993)}]{Dierckx1993}Dierckx, P.  1993, Curve and surface fitting with splines (New York, NY, USA:Oxford University Press, Inc.)

\bibitem[{{Drake} {et~al.}(2009)}]{2009ApJ...696..870D}{Drake}, A.~J., {et~al.} 2009, \apj, 696,870

\bibitem[{{Durbin} \& {Watson}(1951)}]{DurbinWatson}{Durbin}, J. \&  {Watson}, G.~S.  1951, Biometrika, 37, 409

\bibitem[{{Filippenko} {et~al.}(2001)}]{2001ASPC..246..121F}{Filippenko}, A.~V., {Li}, W.~D., {Treffers}, R.~R., \&  {Modjaz}, M.  2001, Small Telescope Astronomy on Global Scales, Vol. 246, ed. W.~P. Chen, C. Lemme, \& B. Paczynski (San Francisco, CA:  ASP), 121

\bibitem[{{Folatelli} {et~al.}(2010)}]{2010AJ....139..120F}{Folatelli}, G., {et~al.} 2010, \aj, 139,120

\bibitem[{{Freedman} \& {Madore}(2010)}]{FreedmanMadore2010}{Freedman}, W.~L. \, \& ~{Madore}, B.~F.  2010, \araa, 48, in press

\bibitem[{{Freedman} {et~al.}(2009)}]{2009ApJ...704.1036F}{Freedman}, W.~L., {et~al.} 2009, \apj, 704, 1036

\bibitem[{{Freedman} {et~al.}(2001)}]{2001ApJ...553...47F}{Freedman}, W.~L., {et~al.} 2001, \apj, 553, 47

\bibitem[{{Frieman} {et~al.} (2008)}]{2008ARA&A..46..385F}{Frieman}, J.~A., {Turner}, M.~S., \&  {Huterer}, D.  2008, \araa, 46, 385

\bibitem[{{Gallagher} {et~al.}(2008)}]{2008ApJ...685..752G}{Gallagher}, J.~S., {et~al.} 2008,\apj, 685, 752

\bibitem[{{Guy} {et~al.}(2007)}]{2007A&A...466...11G}{Guy}, J., {et~al.} 2007, \aap, 466, 11

\bibitem[{{Guy} {et~al.}(2005)}]{2005A&A...443..781G}{Guy}, J., {et~al.} 2005,\aap, 443, 781

\bibitem[{{Hamuy} {et~al.}(2006)}]{2006PASP..118....2H}{Hamuy}, M., {et~al.} 2006, \pasp, 118, 2

\bibitem[{{Hamuy} {et~al.}(1996{\natexlab{a}})}]{1996AJ....112.2391H}{Hamuy}, M., {et~al.} 1996{\natexlab{a}}, \aj, 112, 2391

\bibitem[{{Hamuy} {et~al.}(1996{\natexlab{b}})}]{1996AJ....112.2438H}{Hamuy}, M., {et~al.} 1996{\natexlab{b}}, \aj, 112,2438

\bibitem[Hatano et al.(1998)]{1998ApJ...502..177H} Hatano, K., Branch, D., 
\& Deaton, J.\ 1998, \apj, 502, 177

\bibitem[{{Hicken} {et~al.}(2009)}]{2009ApJ...700.1097H}{Hicken}, M., {et~al.} 2009, \apj, 700, 1097

\bibitem[{{Hsiao} {et~al.}(2007)}]{2007ApJ...663.1187H}{Hsiao}, E.~Y., {et~al.} 2007, \apj,663, 1187

\bibitem[{{Jha} {et~al.}(2007)}]{2007ApJ...659..122J}{Jha}, S., {Riess}, A.~G., \&  {Kirshner}, R.~P.  2007, \apj, 659, 122

\bibitem[{{Johnson}(1963)}]{1963bad..book..204J}{Johnson}, H.~L.  1963, {Photometric Systems} (the University of Chicago Press),204

\bibitem[{{Kelly}(2007)}]{2007ApJ...665.1489K}{Kelly}, B.~C.  2007, \apj, 665, 1489

\bibitem[{{Kessler} {et~al.}(2009)}]{2009ApJS..185...32K}{Kessler}, R., {et~al.} 2009, \apjs, 185,32

\bibitem[{{Krisciunas} {et~al.}(2004)}]{2004ApJ...602L..81K}{Krisciunas}, K., {Phillips}, M.~M., \&  {Suntzeff}, N.~B.  2004, \apjl, 602, L81

\bibitem[{{Lampeitl} {et al.}(2010)}]{2010arXiv1005.4687L} Lampeitl, H., et al.\ 
2010, arXiv:1005.4687 

\bibitem[{{Landolt}(1992)}]{1992AJ....104..340L}{Landolt}, A.~U.  1992, \aj, 104, 340

\bibitem[{{Lira}(1996)}]{1996MsT..........3L}{Lira}, P.  1996, Master's thesis, MS thesis.~Univ.~Chile (1996)

\bibitem[{{Mandel} {et~al.}(2009)}]{2009ApJ...704..629M}{Mandel}, K.~S., {Wood-Vasey}, W.~M., {Friedman}, A.~S., \&  {Kirshner}, R.~P. 2009, \apj, 704, 629

\bibitem[{{Nugent} {et~al.}(2002)}]{2002PASP..114..803N}{Nugent}, P., {Kim}, A., \&  {Perlmutter}, S.  2002, \pasp, 114, 803

\bibitem[{{O'Donnell}(1994)}]{1994ApJ...422..158O}{O'Donnell}, J.~E.  1994, \apj, 422, 158

\bibitem[{{Perlmutter} {et~al.}(1999)}]{1999ApJ...517..565P}{Perlmutter}, S., {et~al.} 1999, \apj, 517, 565

\bibitem[{{Persson} {et~al.}(2004)}]{2004AJ....128.2239P}{Persson}, S.~E., {et~al.} 2004, \aj, 128, 2239

\bibitem[{{Phillips}(1993)}]{1993ApJ...413L.105P}{Phillips}, M.~M.  1993, \apjl, 413, L105

\bibitem[{{Phillips} {et~al.}(1999)}]{1999AJ....118.1766P}{Phillips}, M.~M., {et~al.} 1999, \aj, 118, 1766

\bibitem[{{Pignata} {et al.}(2009)}]{2009RMxAC..35R.317P} Pignata, G., Maza, J., 
Hamuy, M., Antezana, R., 
\& Gonzales, L.\ 2009, Revista Mexicana de Astronomia y Astrofisica Conference Series, 35, 317

\bibitem[{{Prieto} {et~al.}(2006)}]{2006ApJ...647..501P}{Prieto}, J.~L., {Rest}, A., \&  {Suntzeff}, N.~B.  2006, \apj, 647, 501

\bibitem[{{Renka}(1987)}]{1987RenKa}{Renka}, R.~J.  1987, SIAM J. Sci. Stat. Comput., 8, 393

\bibitem[{{Riess} {et~al.}(1998)}]{1998AJ....116.1009R}{Riess}, A.~G., {et~al.} 1998, \aj, 116, 1009

\bibitem[{{Riess} {et~al.}(2009)}]{2009ApJ...699..539R}{Riess}, A.~G., {et~al.} 2009, \apj, 699, 539

\bibitem[{{Riess} {et~al.}(1996)}]{1996ApJ...473...88R}{Riess}, A.~G., {Press}, W.~H., \&  {Kirshner}, R.~P.  1996, \apj, 473, 88

\bibitem[{{Schlegel} {et~al.}(1998)}]{1998ApJ...500..525S}{Schlegel}, D.~J., {Finkbeiner}, D.~P., \&  {Davis}, M.  1998, \apj, 500, 525

\bibitem[{{Schweizer} {et~al.}(2008)}]{2008AJ....136.1482S}{Schweizer}, F., {et~al.} 2008, \aj, 136, 1482

\bibitem[{{Smith} {et~al.}(2002)}]{2002AJ....123.2121S}{Smith}, J.~A., {et~al.} 2002, \aj, 123, 2121

\bibitem[{{Spergel} {et~al.}(2007)}]{2007ApJS..170..377S}{Spergel}, D.~N., {et~al.} 2007, \apjs, 170, 377

\bibitem[{{Stritzinger} {et~al.}(2010)}]{Stritzinger2010}{Stritzinger}, M., {et~al.}  2010, \aj, in press

\bibitem[{{Stritzinger} {et~al.}(2002)}]{2002AJ....124.2100S}{Stritzinger}, M., {et~al.} 2002, \aj,124, 2100

\bibitem[{{Stritzinger} {et~al.}(2005)}]{2005PASP..117..810S}{Stritzinger}, M., {et~al.} 2005, \pasp, 117, 810

\bibitem[{{Sullivan} {et~al.}(2010)}]{2010MNRAS.tmp..755S}{Sullivan}, M., {et~al.} 2010, \mnras, 755

\bibitem[{Thijsse {et~al.}(1998)}]{297968}Thijsse, B.~J., Hollanders, M.~A., \&  Hendrikse, J.  1998, Comput. Phys., 12,393

\bibitem[{{Tripp}(1998)}]{1998A&A...331..815T}{Tripp}, R.  1998, \aap, 331, 815

\bibitem[{{Tripp} \& {Branch}(1999)}]{1999ApJ...525..209T}{Tripp}, R., \&  {Branch}, D.  1999, \apj, 525, 209

\bibitem[{{Tully} {et~al.}(2008)}]{2008ApJ...686.1523T}{Tully}, R.~B., {et~al.} 2008, \apj, 686, 1523

\bibitem[{{Wood-Vasey} {et~al.}(2008)}]{2008ApJ...689..377W}{Wood-Vasey}, W.~M., {et~al.} 2008, \apj, 689, 377

\bibitem[{{Wood-Vasey} {et~al.}(2007)}]{2007ApJ...666..694W}{Wood-Vasey}, W.~M., {et~al.} 2007, \apj, 666, 694

\end{thebibliography}

\clearpage

\clearpage
\clearpage
\begin{deluxetable}{lcclcclr}
\rotate
\tablecolumns{8}
\tablewidth{0pc}
\tablecaption{SN~Ia used for Calibration and Template Generation\label{tab:SNe}}
\tablehead{\colhead{Name} & \colhead{$v_{hel}$} & \colhead{$v_{cmb}$} &
           \colhead{Host} &
           \colhead{$t_0$} & 
           \colhead{$E(B-V)_{MW}$} &  
           \colhead{Template} & \colhead{Notes}\\
           \colhead{SN} & \colhead{${\rm km\cdot s^{-1}}$} & 
           \colhead{${\rm km\cdot s^{-1}}$} & \colhead{} &
           \colhead{days} & \colhead{mag}
           & \colhead{} & \colhead{}\\
           \colhead{(1)} & \colhead{(2)} & \colhead{(3)} & \colhead{(4)} & \colhead{(5)}
           & \colhead{(6)} & \colhead{(7)} & \colhead{(8)}}
\startdata
2004ef & 9289 & 8931 & UGC 12158 & -8.8 & 0.06 & $uBVgri$ &  \\
2004eo & 4706 & 4421 & NGC 6928 & -12.2 & 0.11 & $uBVgriYJH$ &  \\
2004ey & 4733 & 4388 & UGC 11816 & -9.0 & 0.14 & $uBVgri$ & Blue \\
2004gc & 9203 & 9214 & PGC 017176 & 5.8 & 0.21 &  &  \\
2004gs & 7988 & 8249 & MCG +03-22-020 & -3.2 & 0.03 & $uBVgri$ &  \\
2004gu & 13748 & 14069 & FGC 175A & -0.3 & 0.03 &  &  \\
2005A & 5738 & 5502 & NGC 0958 & -3.9 & 0.03 & $uBVgr$ & Red \\
2005M & 6599 & 6891 & NGC 2930 & -8.0 & 0.03 & $uBVgriYJH$ & Blue \\
2005W & 2665 & 2385 & NGC 0691 & -8.0 & 0.07 & $uBVgri$ &  \\
2005ag & 23811 & 24024 & MAPS-NGP O 502 0366176 & -1.6 & 0.04 &  & Blue \\
2005al & 3718 & 3986 & NGC 5304 & -0.7 & 0.05 & $uBVgr$ & Blue \\
2005am & 2368 & 2690 & NGC 2811 & -4.1 & 0.05 & $uBVgri$ & Blue \\
2005be & 10500 & 10673 & NPM1G +16.0412 & 7.5 & 0.03 &  &  \\
2005bg & 6921 & 7247 & MCG +03-31-093 & 1.3 & 0.03 &  &  \\
2005bl & 7213 & 7534 & NGC 4070 & -5.6 & 0.03 &  & Fast \\
2005bo & 4166 & 4504 & NGC 4708 & -0.3 & 0.05 &  &  \\
2005el & 4470 & 4465 & NGC 1819 & -7.3 & 0.11 & $uBVgriYJH$ & Blue \\
2005eq & 8687 & 8505 & MCG -01-09-006 & -3.4 & 0.07 & $BVgrYJH$ &  \\
2005hc & 13772 & 13503 & MCG +00-06-003 & -4.7 & 0.03 & $uBVgr$ & Blue \\
2005iq & 10206 & 9879 & ESO 538- G 013 & -5.4 & 0.02 & $uBVgri$ & Blue \\
2005ir & 22892 & 22570 & SDSS J011643.87+004736.9 & -2.2 & 0.03 &  &  \\
2005kc & 4533 & 4167 & NGC 7311 & -10.7 & 0.13 & $uBVgriYJH$ &  \\
2005ke & 1463 & 1345 & NGC 1371 & -9.0 & 0.03 & $uBVgriYJH$ & Blue,Fast \\
2005ki & 5758 & 6111 & NGC 3332 & -9.7 & 0.03 & $uBVgriYJH$ & Blue \\
2005lu & 9596 & 9388 & ESO 545- G 038 & 9.6 & 0.03 &  &  \\
2005na & 7891 & 8044 & UGC 03634 & -1.9 & 0.08 & $BVgr$ &  \\
2006D & 2556 & 2892 & MRK 1337 & -5.9 & 0.05 & $BVgriYJH$ & Blue \\
2006X & 1571 & 1896 & NGC 4321 & -10.0 & 0.03 & $YJH$ & Red \\
2006ax & 5018 & 5387 & NGC 3663 & -11.7 & 0.05 & $uBVgriYJH$ & Blue \\
2006bh & 3252 & 3148 & NGC 7329 & -4.8 & 0.03 & $uBVgriYJH$ & Blue \\
2006eq & 14840 & 14510 & 2MASX J21283758+0113490 & 5.3 & 0.05 &  &  \\
2006et & 6652 & 6494 & NGC 232 & -7.4 & 0.02 & $uBVgriYJH$ &  \\
2006gt & 13422 & 13093 & 2MASX J00561810-0137327 & -1.3 & 0.04 &  & Fast \\
2006mr & 1760 & 1653 & NGC 1316 & -4.2 & 0.02 & $uBVgriYJH$ & Fast \\
2006py & 17357 & 16993 & SDSS J224142.04-000812.9 & -1.5 & 0.06 &  &  \\
2007af & 1639 & 1887 & NGC 5584 & -9.0 & 0.04 & $uBVgriYJH$ &  \\
\enddata
\tablecomments{Columns: (1) SN~ name; (2) heliocentric radial velocity from the 
NASA/IPAC Extragalactic Database (NED); (3) radial velocity in the frame of
the Cosmic Microwave Background (CMB) from NED; (4) host galaxy name from NED;
(5) time of earliest photometric observation relative to $B$-band maximum;
(6) Milky Way reddening \citep{1998ApJ...500..525S}; (7) list of filters for
which this object was used to generate light-curve templates; (8) Notes: ``Blue"
denotes an unreddened SN~~Ia from F10, ``Red" denotes a highly-reddened SN~Ia
from F10, and ``Fast" denotes a fast-declining (e.g., SN~~1991bg-like) SN~Ia.}
\end{deluxetable}

\clearpage
\begin{deluxetable}{lccccccr}
\tablecolumns{7}
\tablewidth{0pc}
\tablecaption{SN~Ia Light Curve Parameters\label{tab:LCparams}}
\tablehead{\colhead{Name} & \colhead{$\Delta m_{15}\left(B\right)$} 
      & \colhead{$\Delta m_{15}$} &
      \colhead{$t_{max}\left(B\right)$} &
      \colhead{$t_{max}$} & \colhead{$E(B-V)$} & 
      \colhead{$\mu$}\\
      \colhead{SN} & \colhead{mag} & \colhead{mag} & \colhead{days} &
      \colhead{days} & \colhead{mag} & \colhead{mag}\\
      \colhead{(1)} & \colhead{(2)} & \colhead{(3)} & \colhead{(4)} & 
      \colhead{(5)} & \colhead{(6)} & \colhead{(7)}}
\startdata
2004ef & 1.36(0.01) & 1.41(0.01) & 264.41(0.04) & 264.02(0.10) & 0.226(0.005) & 35.37(0.01)\\
2004eo & 1.38(0.05) & 1.33(0.01) & 278.84(0.35) & 278.25(0.12) & 0.131(0.006) & 33.81(0.01)\\
2004ey & 0.93(0.01) & 1.00(0.01) & 304.40(0.10) & 304.07(0.10) & -0.035(0.003) & 33.92(0.01)\\
2004gc & \ldots & 1.08(0.02) & \ldots & 323.55(0.41) & 0.188(0.010) & 35.26(0.02)\\
2004gs & 1.61(0.04) & 1.53(0.01) & 356.49(0.08) & 355.58(0.12) & 0.299(0.005) & 35.47(0.01)\\
2004gu & \ldots & 0.74(0.01) & \ldots & 361.66(0.22) & 0.125(0.008) & 36.32(0.01)\\
2005A & 1.12(0.04) & 1.11(0.01) & 380.33(0.16) & 380.00(0.20) & 1.083(0.011) & 34.46(0.02)\\
2005M & 0.83(0.02) & 0.79(0.01) & 405.86(0.26) & 405.25(0.10) & 0.044(0.004) & 35.00(0.01)\\
2005W & 1.22(0.06) & 1.15(0.02) & 412.34(0.04) & 412.05(0.12) & 0.221(0.004) & 32.73(0.01)\\
2005ag & \ldots & 0.92(0.01) & \ldots & 413.89(0.17) & 0.021(0.005) & 37.54(0.01)\\
2005al & 1.16(0.02) & 1.24(0.01) & 430.28(0.13) & 429.95(0.14) & 0.014(0.004) & 33.95(0.01)\\
2005am & 1.47(0.06) & 1.48(0.01) & 436.93(0.24) & 436.16(0.13) & 0.156(0.004) & 32.33(0.01)\\
2005be & \ldots & 1.49(0.03) & \ldots & 459.91(0.49) & 0.029(0.019) & 35.82(0.02)\\
2005bg & \ldots & 0.99(0.02) & \ldots & 469.80(0.34) & -0.028(0.017) & 35.06(0.03)\\
2005bo & \ldots & 1.30(0.02) & \ldots & 478.55(0.20) & 0.347(0.005) & 33.87(0.01)\\
2005el & 1.34(0.01) & 1.39(0.01) & 647.56(0.03) & 646.65(0.11) & -0.005(0.004) & 33.88(0.01)\\
2005eq & 0.72(0.04) & 0.74(0.01) & 654.24(0.36) & 653.73(0.15) & 0.033(0.004) & 35.33(0.01)\\
2005hc & 0.90(0.01) & 0.85(0.01) & 667.39(0.08) & 666.91(0.12) & 0.019(0.004) & 36.43(0.01)\\
2005iq & 1.25(0.02) & 1.30(0.01) & 688.14(0.08) & 687.48(0.11) & 0.022(0.004) & 35.81(0.01)\\
2005ir & \ldots & 0.86(0.02) & \ldots & 684.35(0.27) & 0.050(0.014) & 37.50(0.03)\\
2005kc & 1.19(0.02) & 1.23(0.01) & 698.06(0.19) & 697.68(0.11) & 0.316(0.005) & 33.82(0.01)\\
2005ke & 1.73(0.01) & 1.81(0.01) & 698.78(0.05) & 698.19(0.10) & \ldots & \ldots\\
2005ki & 1.37(0.01) & 1.37(0.01) & 706.07(0.06) & 705.08(0.11) & 0.020(0.003) & 34.58(0.01)\\
2005lu & \ldots & 0.74(0.04) & \ldots & 710.53(0.57) & 0.215(0.011) & 35.85(0.02)\\
2005na & 0.95(0.04) & 1.07(0.01) & 740.20(0.34) & 740.16(0.15) & -0.018(0.007) & 35.12(0.02)\\
2006D & 1.38(0.01) & 1.47(0.01) & 757.75(0.03) & 757.26(0.10) & 0.204(0.004) & 32.71(0.01)\\
2006X & 1.09(0.02) & 0.84(0.01) & 786.08(0.27) & 785.23(0.15) & 1.345(0.012) & 30.99(0.02)\\
2006ax & 1.00(0.01) & 0.97(0.01) & 827.53(0.12) & 826.85(0.11) & -0.036(0.004) & 34.20(0.01)\\
2006bh & 1.37(0.02) & 1.44(0.01) & 833.55(0.11) & 833.17(0.10) & 0.065(0.003) & 33.27(0.01)\\
2006eq & \ldots & 1.62(0.01) & \ldots & 975.82(0.47) & 0.311(0.011) & 36.57(0.02)\\
2006et & 0.85(0.03) & 0.84(0.01) & 993.55(0.36) & -52005.86(0.13) & 0.183(0.006) & 34.68(0.01)\\
2006gt & \ldots & 1.66(0.01) & \ldots & 1002.45(0.17) & 0.288(0.011) & 36.47(0.02)\\
2006mr & 1.94(0.03) & 2.05(0.01) & 1050.62(0.10) & 1050.35(0.11) & \ldots & \ldots\\
2006py & \ldots & 1.06(0.05) & \ldots & 1070.51(0.36) & 0.094(0.011) & 36.79(0.03)\\
2007af & 1.20(0.03) & 1.11(0.01) & 1174.78(0.14) & -51825.20(0.10) & 0.183(0.003) & 31.90(0.00)\\
\enddata
\tablenotetext{a}{Denotes unreddened SN~Ia from F10.}
\tablenotetext{b}{Denotes highly-reddened SN~Ia from F10.}
\tablenotetext{a}{Classified as a fast-decliner.}
\tablecomments{Columns: (1) SN name; (2) decline-rate parameter measured
directly from $B$-band light curve (see \S \ref{sec:splines}); (3) 
decline-rate parameter derived from SNooPy fits; (4) date of $B$-band 
maximum measured from $B$-band light curve (${\rm JD} - 2453000$); (5) 
date of $B$-band maximum derived from SNooPy fits; (6) color excess
derived from SNooPy fits; (7) distance modulus derived from SNooPy
fits for SNe~Ia with $\Delta m_{15} < 1.7$ and two red SNe~Ia excluded.}

\end{deluxetable}

\begin{deluxetable}{lccccc}
\tablecolumns{6}
\tablewidth{0pc}
\tablecaption{Extrapolation Errors in Light-Curve Parameters\label{tab:extrap-error}}
\tablehead{\colhead{} 
& \multicolumn{5}{c}{Time of Earliest Observation} \\
\colhead{Parameter} & \colhead{0 days} & \colhead{5 days} & \colhead{10 days} 
 & \colhead{15 days} & \colhead{20 days}  \\
}
\startdata
$t_{max}$ & 0.07 & 0.16 & 0.21 & 0.21 & 0.46 \\
$\dm$ & 0.00 & 0.00 & 0.03 & 0.03 & 0.03 \\
$u_{max}$ & 0.00 & 0.03 & 0.05 & 0.06 & 0.06 \\
$B_{max}$ & 0.00 & 0.02 & 0.02 & 0.03 & 0.04 \\
$V_{max}$ & 0.00 & 0.00 & 0.01 & 0.03 & 0.03 \\
$g_{max}$ & 0.00 & 0.01 & 0.01 & 0.01 & 0.02 \\
$r_{max}$ & 0.00 & 0.01 & 0.02 & 0.02 & 0.02 \\
$i_{max}$ & 0.01 & 0.03 & 0.05 & 0.05 & 0.05 \\
$Y_{max}$ & 0.01 & 0.03 & 0.03 & 0.03 & 0.04 \\
$J_{max}$ & 0.02 & 0.04 & 0.05 & 0.05 & 0.05 \\
$H_{max}$ & 0.01 & 0.02 & 0.02 & 0.02 & 0.02 \\
\enddata
\end{deluxetable}

\clearpage
\begin{deluxetable}{llcccccccc}
\rotate
\tablecolumns{10}
\tablewidth{0pc}
\tablecaption{Tripp Calibration Parameters\label{tab:trip-param}}
\tablehead{\colhead{} & \colhead{} & \multicolumn{4}{c}{Red SNe incl.} & 
\multicolumn{4}{c}{Red SNe excl.} \\
\colhead{Filter} & \colhead{Color} & \colhead{$M_X^{YZ}$} & \colhead{$b_X^{YZ}$} & 
\colhead{$R_X^{YZ}$} &
\colhead{$\sigma_{SN}$} & \colhead{$M_X^{YZ}$} & \colhead{$b_X^{YZ}$} & 
\colhead{$R_X^{YZ}$} & \colhead{$\sigma_{SN}$}\\
}\startdata
$u$ & $g-r$ & -18.19 (0.05) & 0.63 (0.15) & 4.1 (0.2) & 0.24 &
-18.22 (0.06) & 0.52 (0.18) & 3.8 (0.5) & 0.23\\
$u$ & $B-V$ & -18.72 (0.05) & 0.48 (0.12) & 4.1 (0.1) & 0.19 &
-18.72 (0.05) & 0.62 (0.16) & 4.0 (0.4) & 0.19\\
$B$ & $B-V$ & -19.12 (0.04) & 0.62 (0.11) & 2.8 (0.1) & 0.15 &
-19.08 (0.04) & 0.42 (0.12) & 3.0 (0.3) & 0.15\\
$V$ & $B-V$ & -19.07 (0.04) & 0.38 (0.09) & 1.8 (0.1) & 0.14 &
-19.08 (0.04) & 0.41 (0.12) & 2.0 (0.3) & 0.15\\
$g$ & $g-r$ & -18.80 (0.04) & 0.43 (0.11) & 2.4 (0.1) & 0.18 &
-18.81 (0.04) & 0.35 (0.14) & 2.4 (0.4) & 0.18\\
$r$ & $g-r$ & -18.80 (0.04) & 0.43 (0.11) & 1.4 (0.1) & 0.17 &
-18.81 (0.04) & 0.36 (0.14) & 1.4 (0.4) & 0.18\\
$i$ & $g-r$ & -18.29 (0.04) & 0.27 (0.11) & 1.1 (0.1) & 0.18 &
-18.28 (0.05) & 0.20 (0.15) & 1.1 (0.4) & 0.19\\
$Y$ & $V-J$ & -18.22 (0.03) & 0.23 (0.08) & 0.3 (0.1) & 0.08 &
-18.26 (0.06) & 0.21 (0.09) & 0.2 (0.1) & 0.04\\
$J$ & $V-J$ & -18.35 (0.04) & 0.33 (0.11) & 0.2 (0.1) & 0.13 &
-18.48 (0.09) & 0.38 (0.13) & -0.1 (0.2) & 0.09\\
$H$ & $V-J$ & -18.19 (0.03) & 0.24 (0.08) & 0.2 (0.1) & 0.07 &
-18.18 (0.08) & 0.20 (0.11) & 0.3 (0.2) & 0.06\\
\enddata
\end{deluxetable}
\clearpage

\begin{deluxetable}{lcccc}
\tablecolumns{5}
\tablewidth{0pc}
\tablecaption{Calibration Parameters for the Reddening Model\label{tab:EBV-global}}
\tablehead{\colhead{Case} & \colhead{$R_V$} & \colhead{$\sigma_{SN}$} &
           \colhead{$\sigma_c$} & \colhead{$\tau$}}
\startdata
Red SNe incl. & 1.40 (0.06) & 0.08 & 0.08 & \ldots \\
Red SNe excl. & 1.95 (0.16) & 0.05 & 0.07 & \ldots \\
Jha Prior + Red SNe incl. & 1.51 (0.07) & 0.09 & 0.10 & 0.26 \\
Jha Prior + Red SNe excl. & 2.64 (0.25) & 0.07 & 0.10 & 0.08 \\
\enddata
\end{deluxetable}

\begin{deluxetable}{lcccccccc}
\tablecolumns{9}
\rotate
\tablewidth{0pc}
\tablecaption{Filter-Specific Calibration Parameters for the Reddening Model\label{tab:EBV-params}}
\tablehead{\colhead{} 
& \multicolumn{2}{c}{Red SNe incl.} & \multicolumn{2}{c}{Red SNe excl.} & \multicolumn{2}{c}{Jha Prior} & \multicolumn{2}{c}{Jha Prior + Red SNe excl.} \\Filter
 & \colhead{$M_0$} & \colhead{$b$} 
 & \colhead{$M_0$} & \colhead{$b$} 
 & \colhead{$M_0$} & \colhead{$b$} 
 & \colhead{$M_0$} & \colhead{$b$} 
}
\startdata
u & -18.62 (0.08) & 0.46 (0.31)  & -18.64 (0.08) & 0.58 (0.32)  & -18.62 (0.16) & 0.25 (0.49)  & -18.52 (0.11) & 0.55 (0.34) \\
B & -19.02 (0.06) & 0.32 (0.24)  & -19.02 (0.06) & 0.32 (0.26)  & -19.03 (0.12) & 0.15 (0.39)  & -18.92 (0.10) & 0.29 (0.29) \\
V & -19.00 (0.04) & 0.32 (0.15)  & -19.01 (0.04) & 0.33 (0.18)  & -19.01 (0.08) & 0.21 (0.25)  & -18.96 (0.07) & 0.30 (0.22) \\
g & -19.07 (0.05) & 0.33 (0.22)  & -19.07 (0.06) & 0.31 (0.24)  & -19.08 (0.11) & 0.18 (0.35)  & -18.98 (0.09) & 0.28 (0.27) \\
r & -18.92 (0.03) & 0.26 (0.12)  & -18.93 (0.04) & 0.26 (0.15)  & -18.94 (0.06) & 0.17 (0.20)  & -18.89 (0.06) & 0.22 (0.19) \\
i & -18.32 (0.02) & 0.11 (0.09)  & -18.35 (0.03) & 0.14 (0.11)  & -18.34 (0.04) & 0.03 (0.13)  & -18.34 (0.05) & 0.09 (0.14) \\
Y & -18.33 (0.02) & 0.10 (0.07)  & -18.35 (0.02) & 0.10 (0.07)  & -18.35 (0.03) & 0.07 (0.09)  & -18.34 (0.03) & 0.06 (0.09) \\
J & -18.43 (0.02) & 0.11 (0.07)  & -18.44 (0.02) & 0.10 (0.07)  & -18.44 (0.03) & 0.09 (0.09)  & -18.44 (0.03) & 0.07 (0.08) \\
H & -18.25 (0.02) & 0.11 (0.07)  & -18.26 (0.02) & 0.12 (0.06)  & -18.26 (0.02) & 0.08 (0.08)  & -18.26 (0.02) & 0.10 (0.07) \\
\enddata
\end{deluxetable}

\end{document}